\documentclass[
  aps,
  prb,
  superscriptaddress,
  floatfix,
  twocolumn,
  secnumarabic,
  amssymb,
  amsmath]{revtex4-1}

\usepackage{dcolumn}
\usepackage{bm}
\usepackage{hyperref}

\usepackage[pdftex]{graphicx}
\usepackage{tabularx}
\usepackage{subfigure}
\usepackage{amssymb}
\usepackage{enumerate}
\usepackage{array}
\usepackage{siunitx}

\usepackage{ifthen}
\usepackage{color}
\usepackage{ulem}             
\definecolor{rot}{rgb}{1,0,0}
\definecolor{blau}{rgb}{0,0,1}
\definecolor{orange}{rgb}{.5,.5,0}
\definecolor{dunkelgruen}{rgb}{.133,0.545,0.133}
\definecolor{dunkelred}{rgb}{.545,0.133,0.133}

\newif\ifcom
\newif\ifdel
\comtrue               %
\delfalse

\begin{document}

\title{Spin Hall magnetoresistance in antiferromagnet/heavy-metal heterostructures}

\author{Johanna~Fischer}
\affiliation{Walther-Mei{\ss}ner-Institut, Bayerische Akademie der Wissenschaften, 85748 Garching, Germany}
\affiliation{Physik-Department, Technische Universit\"{a}t M\"{u}nchen, 85748 Garching, Germany}
\author{Olena Gomonay}
\affiliation{Institut f\"{u}r Physik, Johannes Gutenberg Universit\"{a}t Mainz, 55128 Mainz, Germany}
\author{Richard~Schlitz}
\affiliation{Institut f\"{u}r Festk{\"o}rper- und Materialphysik, Technische Universit\"{a}t Dresden, 01062 Dresden, Germany}
\affiliation{Center for Transport and Devices of Emergent Materials, Technische Universit\"{a}t Dresden, 01062 Dresden, Germany}
\author{Kathrin~Ganzhorn}
\affiliation{Walther-Mei{\ss}ner-Institut, Bayerische Akademie der Wissenschaften, 85748 Garching, Germany}
\affiliation{Physik-Department, Technische Universit\"{a}t M\"{u}nchen, 85748 Garching, Germany}
\author{Nynke~Vlietstra}
\affiliation{Walther-Mei{\ss}ner-Institut, Bayerische Akademie der Wissenschaften, 85748 Garching, Germany}
\affiliation{Physik-Department, Technische Universit\"{a}t M\"{u}nchen, 85748 Garching, Germany}
\author{Matthias~Althammer}
\affiliation{Walther-Mei{\ss}ner-Institut, Bayerische Akademie der Wissenschaften, 85748 Garching, Germany}
\affiliation{Physik-Department, Technische Universit\"{a}t M\"{u}nchen, 85748 Garching, Germany}
\author{Hans~Huebl}
\affiliation{Walther-Mei{\ss}ner-Institut, Bayerische Akademie der Wissenschaften, 85748 Garching, Germany}
\affiliation{Physik-Department, Technische Universit\"{a}t M\"{u}nchen, 85748 Garching, Germany}
\affiliation{Nanosystems Initiative Munich, 80799 M\"{u}nchen, Germany}
\author{Matthias~Opel}
\affiliation{Walther-Mei{\ss}ner-Institut, Bayerische Akademie der Wissenschaften, 85748 Garching, Germany}
\author{Rudolf~Gross}
\affiliation{Walther-Mei{\ss}ner-Institut, Bayerische Akademie der Wissenschaften, 85748 Garching, Germany}
\affiliation{Physik-Department, Technische Universit\"{a}t M\"{u}nchen, 85748 Garching, Germany}
\affiliation{Nanosystems Initiative Munich, 80799 M\"{u}nchen, Germany}
\author{Sebastian~T.B.~Goennenwein}
\affiliation{Institut f\"{u}r Festk{\"o}rper- und Materialphysik, Technische Universit\"{a}t Dresden, 01062 Dresden, Germany}
\affiliation{Center for Transport and Devices of Emergent Materials, Technische Universit\"{a}t Dresden, 01062 Dresden, Germany}
\author{Stephan~Gepr\"{a}gs}
\email[]{stephan.gepraegs@wmi.badw.de}
\affiliation{Walther-Mei{\ss}ner-Institut, Bayerische Akademie der Wissenschaften, 85748 Garching, Germany}

\date{\today}
\begin{abstract}
We investigate the spin Hall magnetoresistance in thin film bilayer heterostructures of the heavy metal Pt and the antiferromagnetic insulator NiO. While rotating an external magnetic field in the easy plane of NiO, we record the longitudinal and the transverse resistivity of the Pt layer and observe an amplitude modulation consistent with the spin Hall magnetoresistance. In comparison to Pt on collinear ferrimagnets, the modulation is phase shifted by 90$^\circ$ and its amplitude strongly increases with the magnitude of the magnetic field. We explain the observed magnetic field-dependence of the spin Hall magnetoresistance in a comprehensive model taking into account magnetic field induced modifications of the domain structure in antiferromagnets. With this generic model we are further able to estimate the strength of the magnetoelastic coupling in antiferromagnets. Our detailed study shows that the spin Hall magnetoresistance is a versatile tool to investigate the magnetic spin structure as well as magnetoelastic effects, even in antiferromagnetic multi-domain materials. 
\end{abstract}

\maketitle

\section{Introduction}
\label{sec:intro}

Spintronic devices integrating ferromagnetic materials and heavy metals (HMs) in multilayer hybrid structures represent well-established basic elements in the field of data storage. For future spintronic applications, \textit{anti}ferromagnetic materials have come into the focus of interest.\cite{Shick:2010,MacDonald:2011,Barthem:2013,Zelezny:2014,Gomonay:2014,Jungwirth:2016,Baltz:2017} They promise robustness against external magnetic field perturbations\cite{Marti:2014,Wadley:2016,Jungwirth:2016} as well as faster magnetization dynamics compared to simple ferromagnets,\cite{Satoh:2014} paving the way to ultrafast information processing.\cite{Duong:2004, Satoh:2010, Kampfrath:2010} Recently, the spin Hall effect (SHE),\cite{Chen:2014,Mendes:2014,Zhang:2014,Ou:2016} the spin Seebeck effect,\cite{Ohnuma:2013,Seki:2015,Rezende:2016,Wu:2016a} and the spin Nernst effect\cite{Cheng:2016} as well as other spin transport phenomena\cite{Wang:2014,Hahn:2014,Moriyama:2015,Lin:2016,Shang:2016,Qaiumzadeh:2017} were discussed in different antiferromagnetic insulators (AFIs) including Cr$_2$O$_3$\cite{Wang:2015} and NiO.\cite{Hou:2017, Lin:2017, Hung:2017} For the integration of such materials in data storage devices, however, a robust detection scheme for their antiferromagnetic magnetization state is required. The spin Hall magnetoresistance (SMR)\cite{Nakayama:2013,Althammer:2013,Vlietstra:2013} could serve as a sensitive probe in this regard. Moreover, the SMR only requires a simple planar metallic electrode on top of the antiferromagnet, making it a promising tool for future applications.

The SMR originates from the interplay of charge and spin currents at the interface between a magnetic insulator (MI) with magnetization $\mathbf{M}$ and a HM with strong spin-orbit coupling. Owing to the spin Hall effect (SHE),\cite{Hirsch:1999} a charge current in the metal leads to an accumulation of a finite spin polarization $\boldsymbol{\sigma}$ at the interface. The exchange of spin angular momentum between $\mathbf{M}$ and $\boldsymbol{\sigma}$ then results in a characteristic dependence of the metal's resistivity on the angle $\angle(\mathbf{M},\boldsymbol{\sigma})$.\cite{Chen:2013} The SMR was first experimentally reported in Y$_{3}$Fe$_{5}$O$_{12}$(YIG)/Pt hybrid structures\cite{Nakayama:2013, Althammer:2013} and theoretically explained by a non-equilibrium proximity effect.\cite{Chen:2013} The validity of the SMR model has been confirmed for YIG/Pt\cite{Nakayama:2013,Althammer:2013,Vlietstra:2013,Hahn:2013,Marmion:2014,Meyer:2014,Aldosary:2016} and other collinear ferrimagnetic insulator/HM systems like YIG/Ta,\cite{Hahn:2013} Fe$_3$O$_4$/Pt,\cite{Althammer:2013} NiFe$_2$O$_4$/Pt,\cite{Althammer:2013} and CoFe$_{2}$O$_{4}$/Pt.\cite{Isasa:2014} Recently, the SMR was used to resolve the orientation of non-collinear magnetic sublattices in canted (Gd$_{3}$Fe$_{5}$O$_{12}$)\cite{Ganzhorn:2016} or spiral (Cu$_{2}$OSeO$_{3}$) ferrimagnets.\cite{Aqeel:2016} In antiferromagnetic thin films, the SMR has been utilized to study the spin transport in exchange-coupled YIG/NiO/Pt bilayer systems.\cite{Shang:2016,Hou:2017, Lin:2017, Hung:2017} Very recently, the bare SMR effect using antiferromagnets was reported for Cr$_2$O$_3$/W\cite{Ji:2017} and bulk NiO/Pt.\cite{Hoogeboom:2017,Baldrati:2017} Furthermore, the SMR response of Cr$_2$O$_3$/Pt, NiO/Pt, and CoO/Pt was recently calculated assuming a single domain antiferromagnet, where the direction of the antiferromagnetic vector is determined by the magnetic anisotropy and the external magnetic field.\cite{Wang:2017}   

In this paper, we systematically investigate the SMR in multidomain antiferromagnetic NiO/Pt bilayer thin films. We use angular-dependent magnetoresistance (ADMR) measurements, rotating the magnetic field in the easy-plane of NiO to measure the SMR amplitude and phase. We find a $90^\circ$ phase shift of the SMR modulation with respect to the SMR observed for collinear ferromagnetic insulators (FMIs).\cite{Althammer:2013} These results demonstrate that the SMR reflects the spin structure of the antiferromagnetically coupled sublattices in NiO. We furthermore observe a pronounced dependence of the SMR amplitude on the applied magnetic field strength. We explain this behavior in the framework of the SMR theory, taking into account magnetic field induced modifications of the antiferromagnetic multidomain state in NiO.

\section{Theory}
\label{sec:theory}

\subsection{Spin Hall magnetoresistance in single-domain antiferromagnetic insulator/heavy-metal bilayers}

\begin{figure}[t]
\includegraphics[width=1.0\columnwidth]{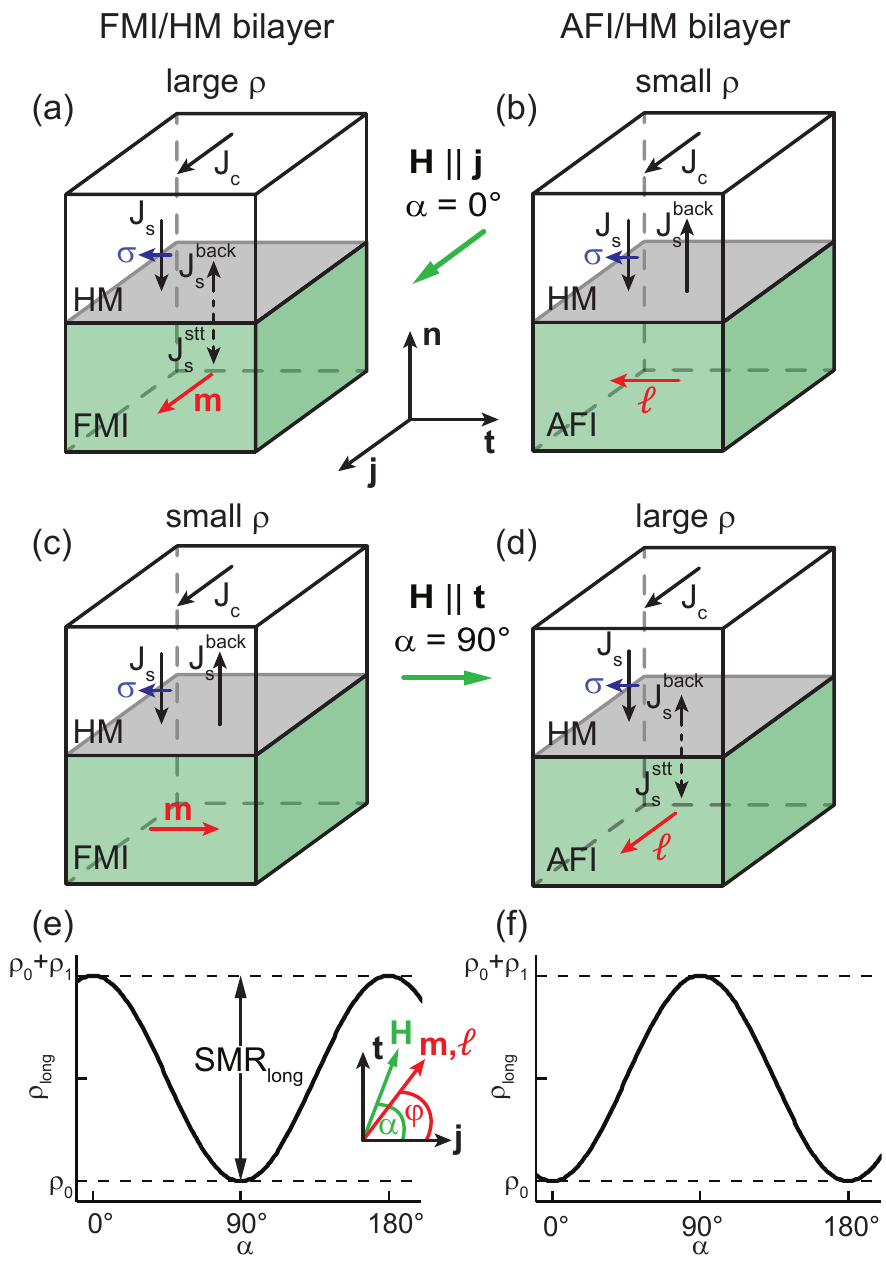}
    \caption{Spin Hall magnetoresistance (SMR) of a single-domain (a,c,e) collinear ferromagnetic insulator/heavy metal (FMI/HM) and (b,d,f) an antiferromagnetic insulator/heavy metal (AFI/HM) bilayer. The SMR is based on an interconversion of charge ($\mathbf{J}_\mathrm{c}$) and spin currents ($\mathbf{J}_\mathrm{s}$) via the spin Hall effect. An increase of the resistivity $\rho_\mathrm{long}$ of the HM is observed, if the spin polarization $\boldsymbol{\sigma}$ of $\mathbf{J}_\mathrm{s}$ is perpendicular to the direction of the order parameter of the magnetic layer (the magnetization direction $\mathbf{m}$ (FMI) or the N\'{e}el vector $\boldsymbol{\ell}$ (AFI)). This leads to a finite spin current $\mathbf{J}_\mathrm{s}^\mathrm{stt}$ in the magnetic layer, which reduces the spin current backflow $\mathbf{J}_\mathrm{s}^\mathrm{back}$ (a,d). For a collinear configuration between $\boldsymbol{\sigma}$ and $\mathbf{m}$ ($\boldsymbol{\ell}$), $\rho_\mathrm{long}$ is approximately given by the normal resistivity of the HM layer (b,c). $\rho_\mathrm{long}$ can be parametrized by the angle $\varphi$ between $\mathbf{m}$ ($\boldsymbol{\ell}$) and the current density direction $\mathbf{j}$. The expected angular-dependence of $\rho_\mathrm{long}$ is sketched in (e,f) as a function of the angle $\alpha$ between the external magnetic field $\mathbf{H}$ and $\mathbf{j}$ for $H$ larger than the anisotropy field (FMI: $\alpha=\varphi$) or the spin-flop field (AFI: $\alpha = 90^\circ + \varphi$).
}
\label{fig:abb1a}
\end{figure}

The spin Hall magnetoresistance (SMR) corresponds to a modulation of the resistance of a HM with strong spin-orbit coupling adjacent to a magnetic insulator (MI) depending on the direction $\mathbf{m}=\mathbf{M}/M$ of its magnetization $\mathbf{M}$.\cite{Chen:2013, Ganzhorn:2016} In such a MI/HM bilayer (see Fig.~\ref{fig:abb1a}), a charge current $\mathbf{J}_\mathrm{c}$ driven through the HM layer induces a spin current $\mathbf{J}_\mathrm{s}$ perpendicular to the spin polarization $\boldsymbol{\sigma}$ and $\mathbf{J}_\mathrm{c}$ via the SHE, creating a local spin accumulation at the MI/HM interface if $\boldsymbol{\sigma}$ is collinear to $\mathbf{m}$. The resulting gradient of the spin accumulation leads to a diffusive spin current backflow $\mathbf{J}_\mathrm{s}^\mathrm{back}$, compensating $\mathbf{J}_\mathrm{s}$. If $\boldsymbol{\sigma}$ is non-collinear to $\mathbf{m}$, a spin transfer torque can be exerted on the magnetic moments resulting in a modification of the spin accumulation and an additional dissipation channel for charge transport in the HM layer and thus an increase of the HM resistivity. \footnote{The magnon accumulation at the MI/HM interface created by the spin accumulation is usually neglected in the description of the SMR.\cite{Cornelissen:2015,Goennenwein:2015}} 

In MIs with $N$ magnetic sublattices, the modulation of the resistivity tensor $\boldsymbol{\rho}$ of the HM layer due to the SMR depends on the directions $\mathbf{m}_p$ with $p=1,\ldots,N$ of the magnetization of each magnetic sublattice.\cite{Ganzhorn:2016} The diagonal component of $\boldsymbol{\rho}$ along the charge current direction $\mathbf{j}$ (see Fig.~\ref{fig:abb1a}), coinciding with the longitudinal resistivity $\rho _{\mathrm{long}}$, is then given by\cite{Ganzhorn:2016}   
\begin{eqnarray}  
\rho_{\mathrm{long}}&=&\rho_{0}+\frac{1}{N}\sum_{p=1}^N \rho_{1,p} \left[ 1 - \left(\mathbf{m}_{p} \cdot \mathbf{t}\right)^{2}\right] \nonumber \\
	& = & \rho_{0}+\frac{1}{N}\sum_{p=1}^N \rho_{1,p}\left[ 1 - m_{p,t}^2\right]
	\; ,
\end{eqnarray}
where $\rho_{0}$ is approximately equal to the normal resistivity of the HM layer\cite{Chen:2013} and $\rho_{1,p}$ represents the SMR coefficient of the $p$th magnetic sublattice with $\rho_{1,p} \ll \rho_0$. $m_{p,t}$ denotes the projection of $\mathbf{m}_{p}$ on $\mathbf{t}$ (perpendicular to $\mathbf{j}$ in the $\mathbf{j}$-$\mathbf{t}$-interface plane, see Fig.~\ref{fig:abb1a}).

Considering a MI/HM bilayer consisting of a FMI with only one magnetic sublattice ($N=1$, see Fig.~\ref{fig:abb1a}(a,c,e)), $\rho_\mathrm{long}$ can be written as
\begin{eqnarray}
  \rho_{\mathrm{long}}&=&\rho_{0}+ \rho_{1}\left[1 - m_{t}^2\right]  
	\; ,
\end{eqnarray}    
with $\rho_1 = \rho_{1,1}$ and $m_{t}=m_{1,t}$. Assuming that the magnetization stays in the $\mathbf{j}$-$\mathbf{t}$ plane, $\rho_{\mathrm{long}}$ depends on the projection $m_{j}$ of the magnetization direction $\mathbf{m}$ on $\mathbf{j}$, i.e. the angle $\varphi$ (see inset of Fig.~\ref{fig:abb1a}(e)), as
\begin{eqnarray}
  \rho_{\mathrm{long}}&=& \; \rho_{0} + \rho_{1} m_{j}^2 \, = \, \rho_{0} + \frac{\rho_{1}}{2} \left( 1+\cos 2\varphi \right)
	\; .
	\label{eq:rholong_FMI}
\end{eqnarray}
This results in a maximum of $\rho_{\mathrm{long}}$ at $\varphi = 0^\circ$ and a minimum of $\rho_{\mathrm{long}}$ at $\varphi = 90^\circ$. From a similar consideration,\cite{Chen:2013,Ganzhorn:2016} the transverse resistivity $\rho_\mathrm{trans}$ is given by
\begin{eqnarray}
  \rho_{\mathrm{trans}}&=& \rho_{3} m_{j} m_{t} \; = \; \frac{\rho_{3}}{2} \sin 2\varphi
	\; ,
	\label{eq:rhotrans_FMI}
\end{eqnarray}
with the transverse SMR coefficient $\rho_3 \ll \rho_0$.\cite{Chen:2013, Althammer:2013} The SMR amplitudes can then be defined as
\begin{eqnarray}
\mathrm{SMR}_\mathrm{long} &=& \frac{\rho_\mathrm{long}(\varphi = 0^\circ)-\rho_\mathrm{long}(\varphi = 90^\circ)}{\rho_\mathrm{long}(\varphi = 90^\circ)} \; = \; \frac{\rho_1}{\rho_0} \nonumber \\
\mathrm{SMR}_\mathrm{trans} &=& \frac{\rho_\mathrm{trans}(\varphi = 45^\circ)-\rho_\mathrm{trans}(\varphi = 135^\circ)}{\rho_\mathrm{long}(\varphi = 90^\circ)} \; = \; \frac{\rho_3}{\rho_0}
\; \; .
\label{eq:SMR_amplitude_FMI}
\end{eqnarray}
With an applied magnetic field $H$ larger than the anisotropy field $H_\mathrm{a}$ of the FMI, $\varphi$ is equal to the angle $\alpha$ between the external magnetic field $\mathbf{H}$ and the current direction $\mathbf{j}$ (see inset of Fig.~\ref{fig:abb1a}(e)). We then expect a $(1+\cos2\alpha)$-dependence of $\rho_{\mathrm{long}}(\alpha)$ as shown in Fig.~\ref{fig:abb1a}(e). 

In an AFI/HM bilayer (see Fig.~\ref{fig:abb1a}(b,d,f)) consisting of an antiferromagnet with two magnetic sublattices ($N=2$), $\rho_\mathrm{long}$ can be written as
\begin{eqnarray}
  \rho_{\mathrm{long}}&=& \rho_{0} + \frac{\rho_{1}}{2} \left[2 - m_{1,t}^2 - m_{2,t}^2\right] 
	\; ,
\end{eqnarray}
with $\rho_{1}=\rho_{1,1}=\rho_{1,2}$ and $m_{1,t}$ ($m_{2,t}$) the projection of $\mathbf{m}_1$ ($\mathbf{m}_2$) on $\mathbf{t}$. Without canting of the sublattice magnetizations, the directions $\mathbf{m}_1$ and $\mathbf{m}_2$ are given by the unit vector (N\'{e}el vector) $\boldsymbol{\ell}=(\mathbf{m}_1-\mathbf{m}_2)/2$ with its projections $\ell_j$ and $\ell_t$ on $\mathbf{j}$ and $\mathbf{t}$, respectively. Assuming that $\boldsymbol{\ell}$ stays in the $\mathbf{j}$-$\mathbf{t}$-plane, we obtain
\begin{eqnarray}
  \rho_{\mathrm{long}}&=& \rho_{0} + \rho_{1} \ell_{j}^2 \; = \; \rho_{0} + \frac{\rho_{1}}{2} \left( 1+\cos 2\varphi \right)  
	\; ,
	\label{eq:rholong_AFI}
  \end{eqnarray}
with the angle $\varphi$ between $\boldsymbol{\ell}$ and $\mathbf{j}$. Similarly, we get
\begin{eqnarray}
  \rho_{\mathrm{trans}}&=& \rho_{3} \ell_j \ell_t \; = \; \frac{\rho_{3}}{2} \sin 2\varphi
	\; .
	\label{eq:rhotrans_AFI}
\end{eqnarray}
While the angular dependence of the resistivity tensor $\boldsymbol{\rho}(\varphi)$ is equal for FMI/HM and AFI/HM bilayers (compare~Eqs.~(\ref{eq:rholong_FMI}), (\ref{eq:rhotrans_FMI}) with (\ref{eq:rholong_AFI}) and (\ref{eq:rhotrans_AFI})) resulting in the same SMR amplitude, it is different with respect to the orientation of the external magnetic field $\boldsymbol{\rho}(\alpha)$. For magnetic fields $H$ larger than the spin-flop field $H_\mathrm{SF}$ of the AFI, $\boldsymbol{\ell}$ is perpendicular to the external magnetic field $\mathbf{H}$, i.e. $\alpha = 90^\circ + \varphi$, resulting in a 90$^\circ$ shift of $\rho_{\mathrm{long}}(\alpha)$ in AFI/HM bilayers with respect to collinear FMI/HM bilayers (see Fig.~\ref{fig:abb1a}(f)). This 90$^\circ$ phase shift is often referred to as a ``negative'' SMR.\cite{Hou:2017,Baldrati:2017} 

\subsection{Spin Hall magnetoresistance in multidomain antiferromagnetic insulator/heavy-metal bilayers}

In real samples, the AFI exhibits different types of domains $k$ with different directions $\boldsymbol{\ell}^{(k)}$. To calculate the total longitudinal and transverse resistivities, we average over the Pt resistance contributions of the HM layer from individual domains $k$: $\rho^{(k)}_{\mathrm{long}}=\rho_{0} + \rho_{1} (l_j^{(k)})^2$ and $\rho^{(k)}_{\mathrm{trans}}=\rho_{3} l^{(k)}_jl^{(k)}_t$ (see Eqs.~(\ref{eq:rholong_AFI}),(\ref{eq:rhotrans_AFI})).\footnote{Since the expected maximum change of the resistivity of the HM layer via the SMR is of the order $10^{-3}$, the difference of the resistivity between a series connection and a more complicated resistance network is expected to be small.} We further neglect any contribution of the antiferromagnetic domain walls, since their influence on the HM resistivity is expected to be small. The averaged total Pt-resistivities taking into account the relative fractions of each domain $\xi_k$ yield   
\begin{eqnarray}\label{eq:averaging_SMR}
	\rho_\mathrm{long}&=&\rho_0+\rho_1\sum_k\xi_k \left(\ell_j^{(k)}\right)^2, \nonumber\\
\rho_\mathrm{trans}&=&\rho_3\sum_k\xi_k \ell^{(k)}_j \ell^{(k)}_t
\, ,
	\end{eqnarray}
with $\sum_k \xi_k = 1$. Therefore, to calculate the SMR amplitude in multidomain AFI/HM heterostructures the knowledge of the antiferromagnetic domain structure in the presence of an external magnetic field $\mathbf{H}$ is required.

\begin{figure*}
\includegraphics[width=1.4\columnwidth]{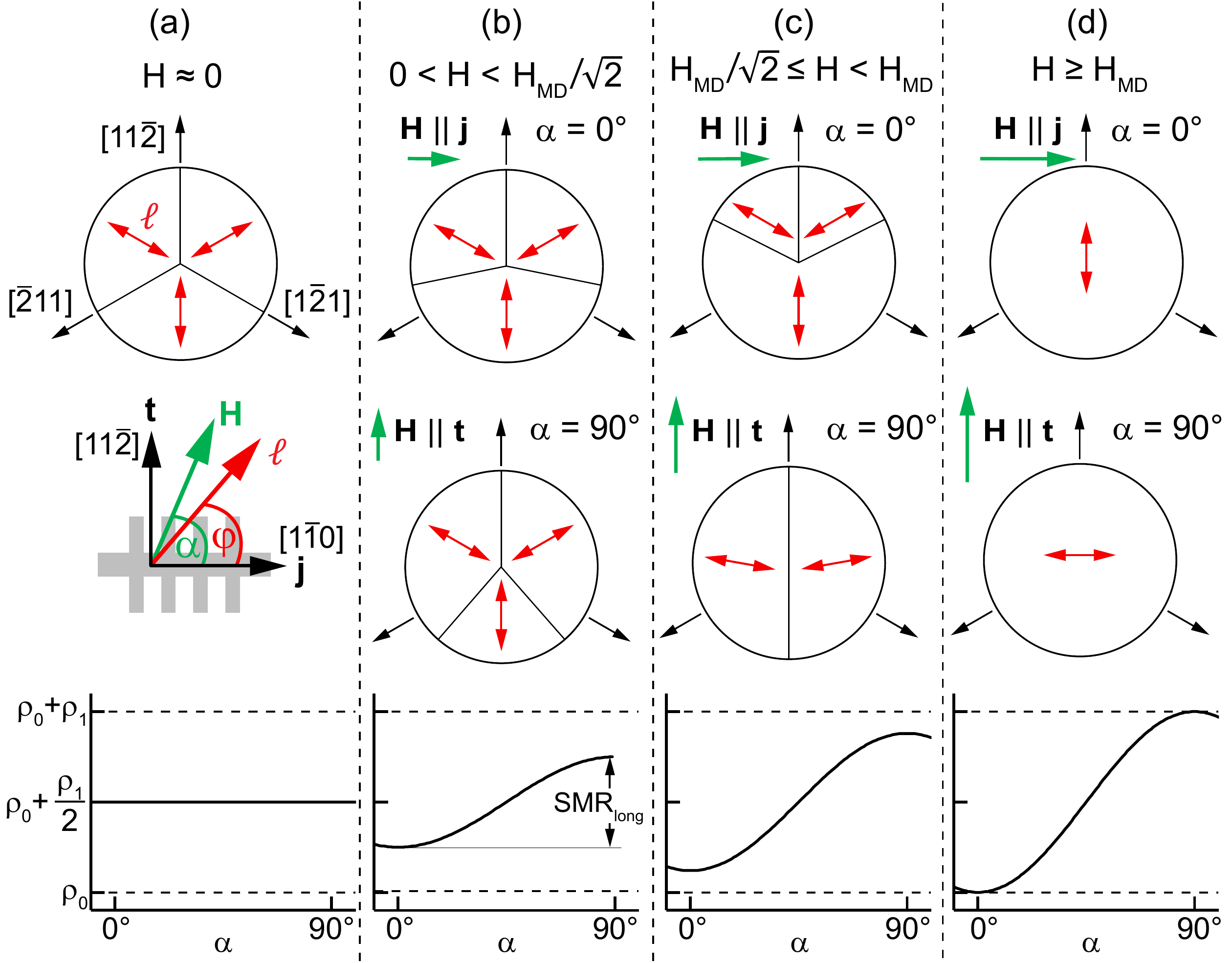}
    \caption{Magnetic configurations in the magnetically easy (111) plane of NiO for an in-plane rotation of the magnetic field $\mathbf{H}$ with $\alpha$ representing the angle between $\mathbf{H}$ and the current density $\mathbf{j}$ for (a) $H \simeq 0$, (b) $0 < H < H_\mathrm{MD}/\sqrt{2}$, (c) $H_\mathrm{MD}/\sqrt{2} \leq H < H_\mathrm{MD}$, and (d) $H \geq H_\mathrm{MD}$ with the monodomainization field $H_{\mathrm{MD}}$. Top: Evolution of the antiferromagnetic multidomain state in NiO with the N\'{e}el vector $\boldsymbol{\ell}^{(k)}$ of each domain $k$ (red double arrows) for an applied magnetic field along $\mathbf{j}$ ($\alpha=0^\circ$). Middle: Same for $\mathbf{H}$ along $\mathbf{t}$ ($\alpha=90^\circ$). Bottom: Expected angular-dependence of the total longitudinal resistivity $\rho_\mathrm{long}$ of a NiO/Pt Hall bar within the SMR theory. The inset shows the orientation of the Pt Hall bar, the magnetic field $\mathbf{H}$, and the N\'{e}el vector $\boldsymbol{\ell}$ with respect to the NiO in-plane directions.
 }
\label{fig:abb1}
\end{figure*}

At zero magnetic field the directions $\boldsymbol{\ell}^{(k)}$ are given by the magnetic anisotropy only. A finite applied magnetic field $\mathbf{H}$ affects the magnetic structure of an antiferromagnet in two ways.\footnote{The external field $\mathbf{H}$ also induces canted magnetization states leading to a finite net magnetization. Due to the usually large exchange field in antiferromagnets, this canting effect can be neglected (exchange approximation), assuming an angle of $180^\circ$ between the sublattice magnetizations, i.e. $\mathbf{m}_1 = -\mathbf{m}_2$, in the considered magnetic field range $H$.} On the one hand, the magnetic field splits the degeneracy of the energetically equivalent domains and creates an effective (ponderomotive) force able to push the domain walls toward the energetically unfavourable domains. On the other hand, the magnetic field induces a coherent rotation of $\boldsymbol{\ell}^{(k)}$ of an individual antiferromagnetic domain $k$ until $\boldsymbol{\ell}^{(k)}$ is perpendicular to $\mathbf{H}$ due to the competition between the magnetic anisotropy energy (favouring the alignment of $\boldsymbol{\ell}$ along the easy axis) and the Zeeman energy (acting to align $\boldsymbol{\ell}$ perpendicular to $\mathbf{H}$).  Which of these mechanisms dominates, depends on the properties of the domain walls. If they are strongly pinned by defects and cannot move under an external magnetic field, the domain structure can be considered as fixed and the magnetic field causes only a coherent rotation of $\boldsymbol{\ell}^{(k)}$. In the opposite case of movable domain walls, the magnetic field modifies the spatial antiferromagnetic domain distribution. This process needs less energy compared to the coherent rotation of $\boldsymbol{\ell}^{(k)}$ within a single domain, as the domain wall motion involves rotation of spins mainly within the domain wall region. Thus, similar to ferromagnets, we can assume that at low magnetic field magnitudes domain redistribution is dominating, while coherent rotation starts at higher magnetic field values when most of the unfavourable domains are removed. This process is schematically shown in Figs.~\ref{fig:abb1}(b-d) for the three-domain case in the easy plane of NiO. 
 
While the external magnetic field triggers domain redistribution, another mechanism, based on magnetoelastic interactions, is responsible for restoring the domain structure after the magnetic field is removed. In contrast to ferromagnets, where the equilibrium domain structure originates from the magnetic dipole-dipole interactions, antiferromagnets show a vanishingly small macroscopic magnetization, which excludes reasonable demagnetization effects. However, antiferromagnets with a pronounced magnetoelastic coupling are subject to strain release effects (destressing effects),\cite{Gomonay:2002} which is an elastic analog to the demagnetization phenomenon in ferromagnets.

In the following we focus on the physics of the destressing effects for the experimentally relevant case of an antiferromagnetic thin film grown on a non-magnetic substrate. The antiferromagnetic ordering is accompanied by the appearance of spontaneous magnetoelastic strains $\hat{u}^{(k)}=\lambda \boldsymbol{\ell}^{(k)}\otimes\boldsymbol{\ell}^{(k)}/G$, where the constant $\lambda$ describes the magnetoelastic coupling, $G$ is the shear modulus, and $\otimes$ denotes the dyadic product. Thus deformed, compared to the non-magnetic state, regions of an antiferromagnet can be treated as magnetoelastic dipoles which, similar to magnetic dipoles, create long-range fields of elastic nature (stresses) in both the magnetic layer and the non-magnetic substrate. Intuitively, in the absence of the magnetic field, the magnetoelastic contribution to the total energy of the sample is minimal in a multidomain state with zero average strain $\langle \hat u\rangle=\sum \xi_k\hat{u}^{(k)}$, since a multidomain state minimizes the total energy. As $\hat{u}$ is related to $\boldsymbol{\ell}$, this condition is also equivalent to zero average of $\langle \boldsymbol{\ell}\otimes\boldsymbol{\ell}\rangle =\sum \xi_k\boldsymbol{\ell}^{(k)}\otimes\boldsymbol{\ell}^{(k)}$.

To calculate the parameters of the multidomain state in the AFI, we introduce the so-called destressing energy density in analogy with the demagnetization energy in ferromagnets representing the main contribution of the energy of magnetoelastic dipoles (for details see Refs.~\onlinecite{Gomonay:2002,Gomonay:2004}):
\begin{equation}
	 E_\mathrm{dest}=\frac{1}{2}H_\mathrm{dest}M\left[\langle \ell_t^2-\ell_j^2\rangle^2+4\langle \ell_t \ell_j\rangle^2\right],
	\label{eq:Edest}
\end{equation}    
where $H_\mathrm{dest}$ is the value of the destressing field, which depends on the properties of the substrate and the interface with the antiferromagnetic layer, $\langle\ldots\rangle$ denotes the mean average over the domain structure, and $M$ is the sublattice magnetization. As obvious from Eq.~(\ref{eq:Edest}), $E_\mathrm{dest}$ is a function of two sets of variables: the N\'{e}el vector $\boldsymbol{\ell}^{(k)}$ inside domains as well as the domain fractions $\xi_k$. Using the coordinate system defined in Fig.~\ref{fig:abb1a}, we can rewrite the destressing energy density as
\begin{equation}\label{eq_destressing_AF_angle}
E_{\mathrm{dest}}= \dfrac{1}{2} H_{\mathrm{dest}} M [ \langle \cos 2 \varphi^{(k)} \rangle ^2 + \langle \sin 2 \varphi^{(k)} \rangle ^2 ] 
\; \; ,
\end{equation} 
where $\varphi^{(k)}$ is the angle between $\boldsymbol{\ell}^{(k)}$ and $\mathbf{j}$.

The domain structure in the presence of an external magnetic field $\mathbf{H}$ can be calculated taking into account the Zeeman energy density with the exchange field $H_\mathrm{ex}$\cite{Uchida:1967} 
\begin{eqnarray}\label{eq:Zeeman_energy}
E_\mathrm{Zee}&=&\frac{M}{4H_\mathrm{ex}}\left[(H_t^2-H_j^2)\langle \ell_t^2- \ell_j^2\rangle+4H_tH_j\langle \ell_t \ell_j\rangle\right] \nonumber \\
	            &=&\dfrac{M H^{2}}{2H_\mathrm{ex}} \langle \cos^{2} (\varphi^{(k)} - \alpha)\rangle -  \dfrac{1}{4} \dfrac{M H^2}{H_\mathrm{ex}}
\end{eqnarray}
with $\alpha$ being the angle between the magnetic field $\mathbf{H}$ and the current direction $\mathbf{j}$ (see Fig.~\ref{fig:abb1a}), as well as the magnetic anisotropy energy density\cite{Uchida:1967}
\begin{equation}\label{eq_anisotropy_AF}
E_{\mathrm{A}}= \dfrac{1}{6}M H_{\mathrm{a}} \langle \cos 6\varphi^{(k)} \rangle
\; \;
\end{equation} 
averaged over the domain structure.

The equilibrium domain structure is then obtained by minimizing the total energy density 
\begin{equation}\label{eq:total_energy}
E_\mathrm{tot}=E_\mathrm{dest}+E_\mathrm{A}+E_\mathrm{Zee}
\; 
\end{equation}    
with respect to $\boldsymbol{\ell}^{(k)}$ and $\xi_k$.

Equations~(\ref{eq:Edest}) and (\ref{eq:Zeeman_energy}) reveal that formally the effects of the destressing and the magnetic field on an antiferromagnet are equivalent. This means that in general the destressing field can fully or partially screen the external magnetic field, in analogy with screening of the external magnetic field due to demagnetization effects in ferromagnets. 

We apply the general theory of equilibrium domain structure to the particular case of NiO/Pt bilayers, where NiO represents a prototypical biaxial AFI with a N\'{e}el temperature of 523\,K,\cite{Srinivasan:1984} crystallizing in a simple sodium chloride structure. Below the N\'{e}el temperature, the Ni$^{2+}$ spins align ferromagnetically along the original cubic $\langle11\overline{2}\rangle$ directions within the \{111\} planes and antiferromagnetically between neighboring \{111\} planes \cite{Roth:1958,Hutchings:1972} thus forming two magnetic sublattices with average directions $\mathbf{m}_{1,2}$. In the absence of an external magnetic field, $\mathbf{m}_{1,2}$ and thus the N\'{e}el vector $\boldsymbol{\ell}$ are aligned along the $\langle 11\overline{2}\rangle $ easy axes, resulting in three types $(k=1,2,3)$ of physically distinguishable antiferromagnetic domains  (cf.~Fig.~\ref{fig:abb1}(a)).\footnote{The three physically distinguishable antiferromagnetic domains result from the threefold symmetry of the NiO(111) plane, i.e. $\mathbf{m}_1$ and $\mathbf{m}_2$ are aligned either parallel or antiparallel to one of the three $[11\overline{2}]$ easy axes (cf. double arrows in Fig.~\ref{fig:abb1}.} 

This domain structure can be described by two independent variables $\xi_k$, as $\sum_{k}\xi_k=1$. These two additional degrees of freedom are enough to fully compensate the effect of two independent combinations of the magnetic field components, $H_t^2-H_j^2$  and $H_tH_j$.  As in this case the effective magnetic field is zero, the orientation $\boldsymbol{\ell}^{(k)}$ within the domain $k$ is defined solely by the magnetic anisotropy and coincides with one of the easy axes  (see Fig.~\ref{fig:abb1}(a)). Hence, the average values  $\langle \ell_t^2- \ell_j^2\rangle$ and $\langle \ell_t \ell_j\rangle$ are proportional to two independent linear combinations of $\xi_k$ and can be treated as independent variables. Minimization of $E_\mathrm{tot}$ then gives the fraction of the energetically unfavourable domains $\xi_\mathrm{unfav}$ and the fraction of the most energetically favourable domains $\xi_\mathrm{fav}$ as a function of the magnetic field orientation $\alpha$ as well as of the magnetic field magnitude $H$ as ($\alpha \le |30^\circ|$)
\begin{eqnarray}\label{eq:frac_3domains}
\xi_\mathrm{unfav,\pm}&=&\dfrac{1}{3} \left[ 1 - \frac{H^2}{4 H_\mathrm{ex} H_\mathrm{dest}} \left( \cos2\alpha \pm \sqrt{3} \sin 2\alpha \right) \right] \; , \nonumber \\
\xi_\mathrm{fav}&=&\dfrac{1}{3} \left[ 1 + \frac{H^2}{2 H_\mathrm{ex} H_\mathrm{dest}} \cos2\alpha \right]
\; .
\end{eqnarray}
From the condition $\xi_\mathrm{unfav}=0$, we distinguish two cases: For $H< \sqrt{2H_\mathrm{dest}H_\mathrm{ex}}$, the sample contains all three types of domains for any $\alpha$ (cf.~Fig.~\ref{fig:abb1}(b)), and in the field range $\sqrt{2H_\mathrm{dest}H_\mathrm{ex}}\le H<2\sqrt{H_\mathrm{dest}H_\mathrm{ex}}$ the sample exhibits two domains for certain values of $\alpha$ and three domains for others (cf.~Fig.~\ref{fig:abb1}(c)). For magnetic field magnitudes $H$ larger than $2 \sqrt{H_\mathrm{dest}H_\mathrm{ex}}$, NiO reaches a single domain state for any $\alpha$ (cf.~Fig.~\ref{fig:abb1}(d)). We therefore identify $2\sqrt{H_\mathrm{dest} H_\mathrm{ex}}$ as the monodomainization field $H_\mathrm{MD}$.

By substituting the obtained domain fractions into Eq.~(\ref{eq:averaging_SMR}), we obtain 
\begin{eqnarray}\label{eq:SMR_3domains}
	\rho_\mathrm{long}&=&\rho_0 + \frac{\rho_1}{2}\left(1-\frac{H^2}{H_\mathrm{MD}^2}\cos2\alpha\right)	\; , \nonumber \\
  \rho_\mathrm{trans}&=&-\frac{\rho_3}{2}\frac{H^2}{H^2_\mathrm{MD}}\sin2\alpha
	\; 
\end{eqnarray}
in the three domain case, i.e. in the field range $0 < H < H_\mathrm{MD}/\sqrt{2}$. The longitudinal resistivity $\rho_\mathrm{long}$ thus oscillates around $\rho_0+\rho_1/2$ as schematically shown in Fig.~\ref{fig:abb1}. The SMR amplitudes are then given by  
\begin{eqnarray}\label{eq:SMR_amplitudes}
\mathrm{SMR}_\mathrm{long} &=& \frac{\rho_1 H^2/H^2_\mathrm{MD}}{\rho_0+\rho_1/2\left(1-H^2/H^2_\mathrm{MD} \right)} \; \approx \; \frac{\rho_1}{\rho_0}\frac{H^2}{H^2_\mathrm{MD}} \; , \nonumber \\
\mathrm{SMR}_\mathrm{trans} &=& \frac{\rho_3 H^2/H^2_\mathrm{MD}}{\rho_0+\rho_1/2\left(1-H^2/H^2_\mathrm{MD} \right)} \; \approx \; \frac{\rho_3}{\rho_0} \frac{H^2}{H^2_\mathrm{MD}}
\; \; .
\end{eqnarray}
Therefore, the amplitudes increase quadratically with the external magnetic field magnitude $H$.
 
At higher magnetic field magnitudes $H_\mathrm{MD}/\sqrt{2} \leq H < H_\mathrm{MD}$, in a two-domain state, only one of the three $\xi_k$ is an independent variable. However, the destressing field still partially compensates the magnetic field, so that 
\begin{equation}\label{eq_average_2domains}
 \langle \ell_t \ell_j\rangle=\frac{H_tH_j}{4H_\mathrm{dest}H_\mathrm{ex}}
\; .
\end{equation} 
In this case, the effective field $H_\mathrm{eff}=H_t^2-H_j^2-H_\mathrm{dest}H_\mathrm{ex}\langle \ell_t^2-\ell_j^2\rangle$ is finite and directed either along $\mathbf{t}$ or along $\mathbf{j}$, causing a coherent rotation of $\boldsymbol{\ell}^{(k)}$ in both domains (see Fig.~\ref{fig:abb1}(c)). To get the equilibrium orientation of $\boldsymbol{\ell}^{(k)}$ with $k=1,2$, the total energy density $E_\mathrm{tot}$ should be minimized taking into account the magnetic anisotropy energy. However, in assumption that $H_\mathrm{A}\ll H_\mathrm{dest}$, which is equal to the assumption that the spin flop field $H_\mathrm{SF}$ is much lower than the monodomainization field $H_\mathrm{MD}$, the anisotropy energy can be neglected and Eqs.~(\ref{eq:SMR_3domains}) and (\ref{eq:SMR_amplitudes}) also become valid in the magnetic field range $H_\mathrm{MD}/\sqrt{2} \le H < H_\mathrm{MD}$, i.e. in the two-domain state.

At even higher magnetic field magnitudes $H \geq  H_\mathrm{MD}$, $\boldsymbol{\ell}^{(k)}$ in each domain $k$ is perpendicular to the effective magnetic field. Thus, the difference between the domains disappears and the sample reaches a single-domain state, where the orientation of $\boldsymbol{\ell}$ is now given mainly by the Zeeman energy, i.e. the external magnetic field $\mathbf{H}$, resulting in a coherent rotation of $\boldsymbol{\ell}$ (cf.~Fig.~\ref{fig:abb1}(d)).
The longitudinal and transverse resistivities are now given by Eqs.~(\ref{eq:rholong_AFI}), (\ref{eq:rhotrans_AFI}) leading to a saturation of the SMR amplitudes $\mathrm{SMR}_\mathrm{long}=\rho_1/\rho_0$ and $\mathrm{SMR}_\mathrm{trans}=\rho_3/\rho_0$, independent of the magnetic field magnitude $H$.

\begin{figure}[t]
\includegraphics[width=1.0\columnwidth]{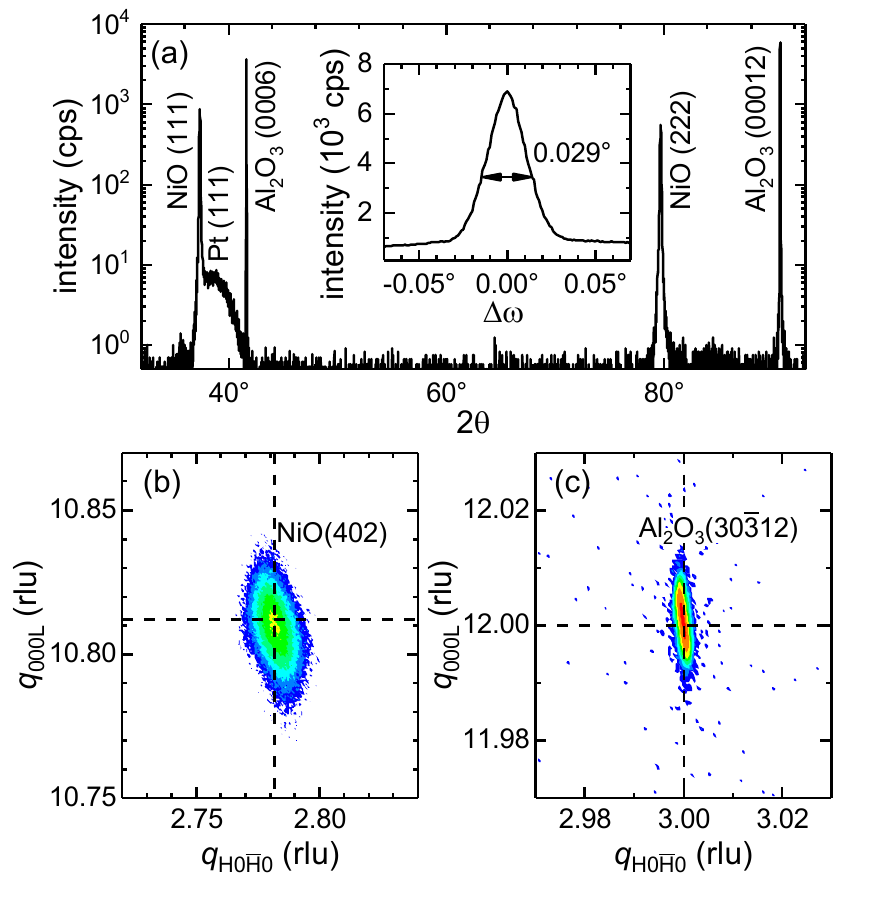}
    \caption{Structural properties of the investigated NiO/Pt heterostructure fabricated on a (0001)-oriented Al$_2$O$_3$ substrate. (a) 2$\theta$-$\omega$-scan along the [0001]-direction of Al$_2$O$_3$. The inset shows the rocking curve around the NiO(111) reflection and the derived full width at half maximum value. (b), (c) Reciprocal space mappings around the NiO(402) and the Al$_2$O$_3$$(3\,0\,\overline{3}\,12)$ reflections. The reciprocal lattice units (rlu) are related to the Al$_2$O$_3$$(3\,0\,\overline{3}\,12)$ substrate reflection.
    }
 \label{fig:abb2a}
\end{figure}

\section{Experiment}
\label{sec:Exp}

\subsection{Sample fabrication}

To corroborate this model we performed systematic ADMR measurements on NiO/Pt bilayer samples. We first fabricated a (111)-oriented NiO thin film on a single crystalline, (0001)-oriented Al$_{2}$O$_{3}$ substrate at $380^\circ$C in an oxygen atmosphere of $10\,\mu$bar via pulsed laser deposition monitored by \textit{in-situ} reflection high energy electron diffraction. Subsequently, the NiO thin film was covered by a thin Pt layer by electron-beam evaporation $\textit{in situ}$ without breaking the vacuum. X-ray diffraction measurements shown in Fig.~\ref{fig:abb2a}(a) reveal a high structural quality of the NiO thin films demonstrated by the full width at half maximum of the rocking curve around the NiO(111) reflection of only $0.029^\circ$. The in-plane orientation and the strain state of NiO was investigated by reciprocal space mappings around the NiO(402) and the Al$_2$O$_3$$(3\,0\,\overline{3}\,12)$ reflections (see Fig.~\ref{fig:abb2a}(b,c)). These measurements reveal the epitaxial relations $[111]_\mathrm{NiO} || [0001]_{\mathrm{Al}_2\mathrm{O}_3}$ and $[1\overline{1}0]_\mathrm{NiO} || [10\overline{1}0]_{\mathrm{Al}_2\mathrm{O}_3}$ of the NiO thin film with respect to the Al$_2$O$_3$ substrate. In addition, a lattice constant of $a_{100}=0.419$\,nm has been derived. This value is close to the bulk lattice constant of NiO ($a=0.4177$\,nm),\cite{Bartel:1971} indicating a nearly fully relaxed strain state of NiO on Al$_2$O$_3$. Furthermore, a low surface roughness below 0.8\,nm (rms value) is confirmed by x-ray reflectometry as well as atomic force microscopy. In the following, we discuss a NiO/Pt thin film bilayer with a thickness of the Pt layer of $t_\mathrm{Pt}=3.5$\,nm and the NiO thin film of $t_\mathrm{NiO}=120$\,nm.

\begin{figure}[t]
\includegraphics[width=0.65\columnwidth]{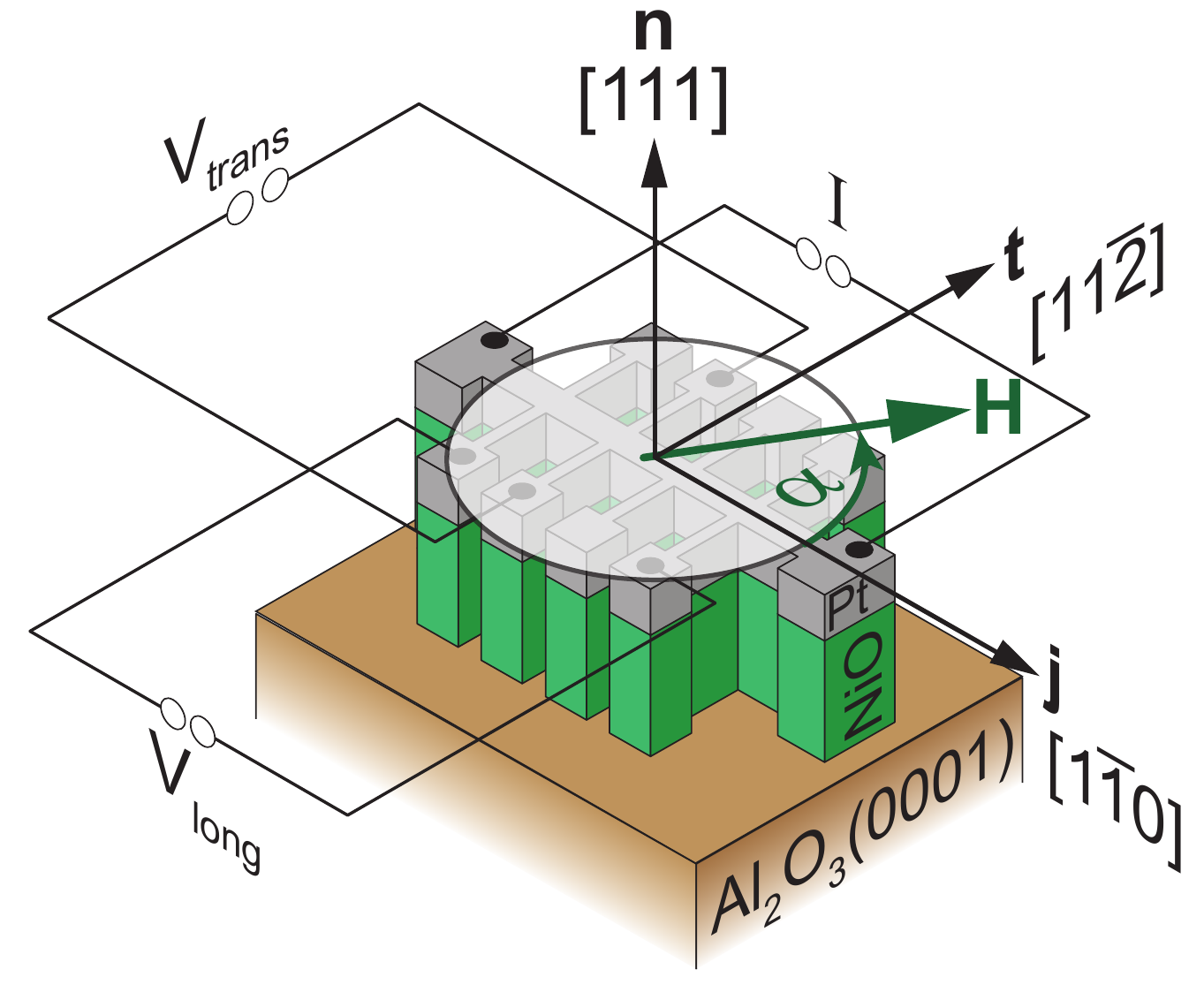}
 \caption{Schematic drawing of the NiO/Pt Hall bar mesa structure with the coordinate system $\mathbf{j}$, $\mathbf{t}$, and $\mathbf{n}$ defined along the crystallographic directions $[1\overline{1}0]$, $[11\overline{2}]$, and $[111]$ of the NiO thin film, respectively. In addition, the measurement scheme used for the magnetotransport measurements with the applied current $I$, the measured longitudinal voltage $V_\mathrm{long}$, and the transverse voltage $V_\mathrm{trans}$ is illustrated. In the NiO(111) plane, the direction of the magnetic field $\mathbf{H}$ is defined by $\alpha$ (green) with respect to the current direction $\mathbf{j}$. $\mathbf{H}$ is rotated counterclockwise.}
\label{fig:abb2b}
\end{figure}

\subsection{Magnetotransport measurements}

For magnetotransport measurements, the sample is patterned into a Hall bar mesa structure via optical lithography and Ar ion milling (see Fig.~\ref{fig:abb2b}). The longitudinal ($\rho_\mathrm{long}$) and transverse resistivities ($\rho_\mathrm{trans}$) are calculated from the longitudinal and the transverse voltages $V_{\mathrm{long}}$ and $V_{\mathrm{trans}}$, measured with a standard four-probe technique using a dc current of $100\,\mu$A and a current-reversal method.\cite{Ganzhorn:2016} We perform ADMR measurements by rotating an externally applied magnetic field of constant magnitude in the (111)-plane of the NiO film as well as sweeping the magnetic field at fixed orientation with respect to the crystallographic axes of NiO at 300\,K.

\begin{figure}[t]
\includegraphics[width=1.0\columnwidth]{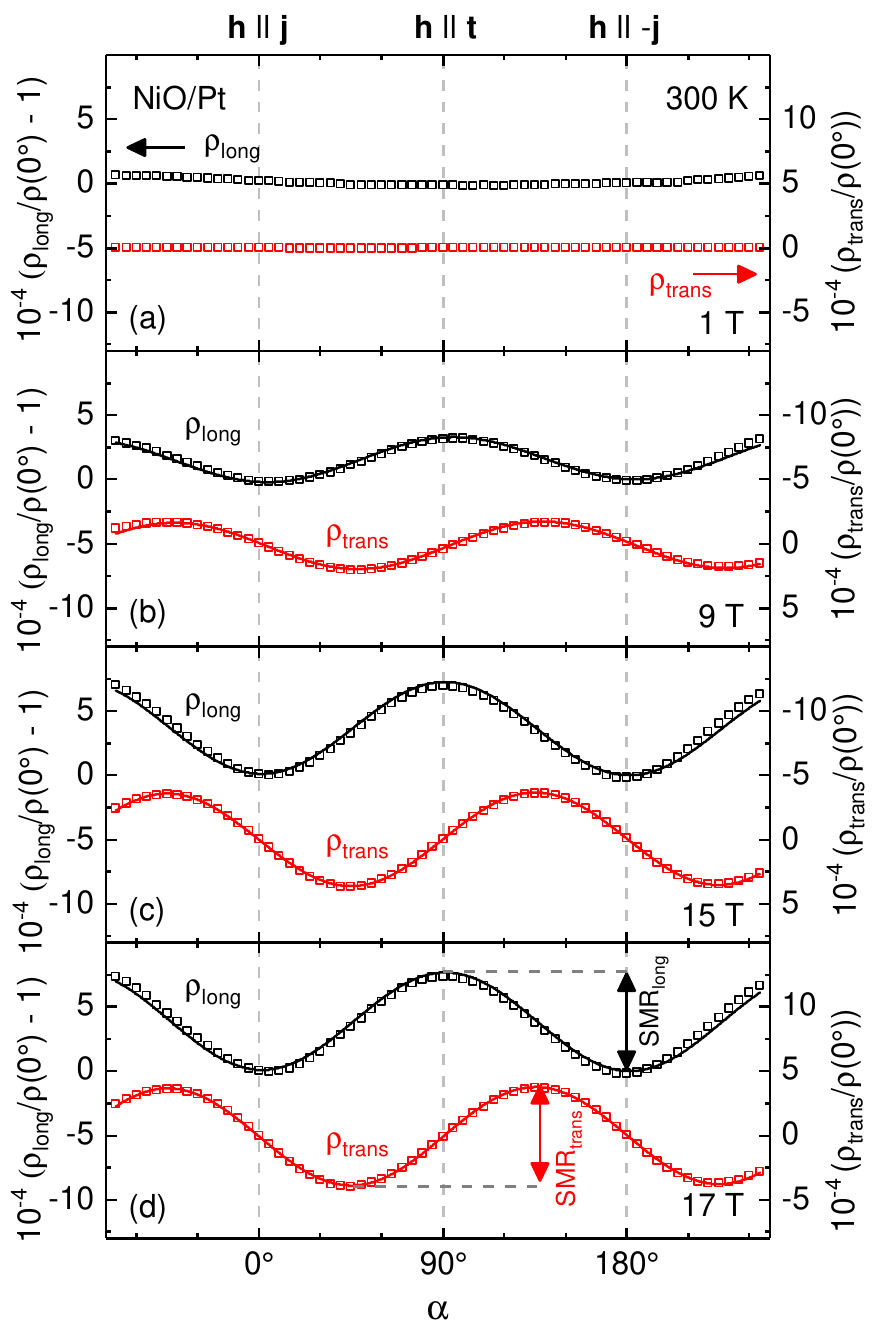}
 \caption{Angular dependent magnetoresistance of a NiO(111)/Pt thin film heterostructure, measured at 300\,K with in-plane external magnetic field magnitudes of (a) 1\,T, (b) 9\,T, (c) 15\,T, (d) 17\,T. Normalized longitudinal resistivity $\rho_\mathrm{long}$ (black symbols, left axis) and transverse resistivity $\rho_\mathrm{trans}$ (red symbols, right axis) as a function of the magnetic field orientation $\alpha$. The lines are fit to the data using $\cos2\alpha$ and $\sin2\alpha$ functions (cf.~Eqs.~(\ref{eq:SMR_3domains})).}
\label{fig:abb2}
\end{figure}

The data obtained from ADMR measurements in different magnetic field magnitudes are shown in Fig.~\ref{fig:abb2}. The predicted $-\cos2\alpha$ dependence of $\rho_\mathrm{long}$ as well as the $-\sin2\alpha$-dependence of $\rho_\mathrm{trans}$ with increasing amplitudes as a function of the applied magnetic field strength (see Eqs.~(\ref{eq:SMR_3domains})) are clearly observed for $\mu_{0}H > 1$\,T. The angular dependence of the resistivities is consistent with the model introduced above for AFI/HM bilayers, i.e.~showing a minimum of $\rho_\mathrm{long}$ at $\alpha = 0^\circ$ and a maximum at $\alpha = 90^\circ$ and being shifted by $90^\circ$ with respect to previous experiments in Pt on collinear ferrimagnets.\cite{Althammer:2013} This provides clear evidence that we are indeed sensitive to $\boldsymbol{\ell}$ (or $\ell_{j}$ and $\ell_{t}$) in the antiferromagnetic NiO as discussed above. The $90^\circ$ phase shift is further consistent with recent experiments in Pt on canted ferrimagnets, where the same shift in the angular dependence is evident close to the compensation temperature,\cite{Ganzhorn:2016} and experiments in YIG/NiO/Pt heterostructures\cite{Shang:2016,Hou:2017, Lin:2017, Hung:2017} as well as NiO/Pt bilayers.\cite{Hoogeboom:2017, Baldrati:2017} For $\mu_{0}H \leq 1$\,T the external magnetic field magnitude $H$ is much smaller than $H_\mathrm{MD}$ resulting in hardly detectable amplitudes of the longitudinal and transverse resistivity variations, respectively. To evaluate the field dependence of the modulation of $\rho_\mathrm{long}$ and $\rho_\mathrm{trans}$ as well as the SMR amplitudes $\mathrm{SMR}_\mathrm{long}$ and $\mathrm{SMR}_\mathrm{trans}$, we fit our data according to Eqs.~(\ref{eq:SMR_3domains}) using $\cos2\alpha$ and $\sin2\alpha$ functions, respectively. The fits are shown as solid lines in Fig.~\ref{fig:abb2}(b-d).

\begin{figure}[t]
\includegraphics[width=1.0\columnwidth]{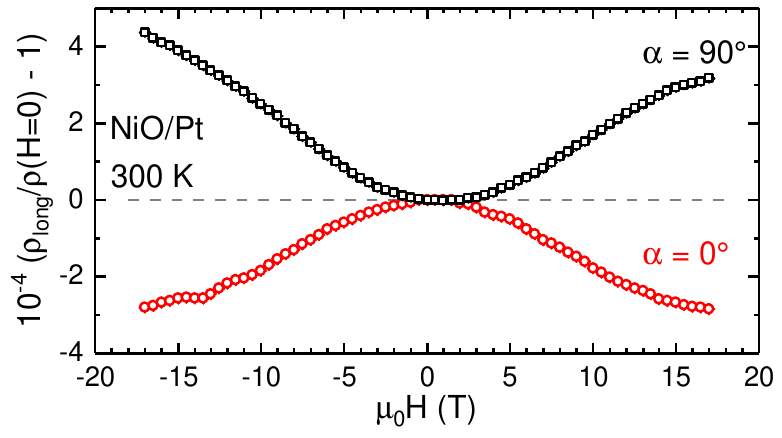}
    \caption{Field dependent longitudinal resistivity $\rho_\mathrm{long}(H)$ of the NiO(111)/Pt bilayer normalized to $\rho_\mathrm{long}(H=0)$ measured at 300\,K for $\alpha=90^\circ$ ($\mathbf{H}\parallel\mathbf{t}$, black symbols) and $\alpha=0^\circ$ ($\mathbf{H}\parallel\mathbf{j}$, red symbols).
    }
 \label{fig:abb3}
\end{figure}

To confirm the magnetic field dependence of $\rho_\mathrm{long}$, we additionally performed field-dependent magnetotransport measurements, sweeping $\mu_0H$ from -17\,T to +17\,T at a fixed orientation $\alpha$. We normalize the data to $\rho(H=0)$ (cf.~Fig.~\ref{fig:abb3}). The slight asymmetry of the signal for $+H$ and $-H$ is caused by variations of the temperature during the field sweeps. For $\alpha=0^\circ$ ($\mathbf{H}\parallel\mathbf{j}$, red symbols) the resistivity decreases with increasing $H$, as the relative fraction of the domain with $\boldsymbol{\ell} \parallel \mathbf{t}$ increases with increasing field (cf.~Fig.~1). According to Eq.~(\ref{eq:averaging_SMR}), this leads to a decrease of $\rho_\mathrm{long}$. For $\alpha=90^\circ$ ($\mathbf{H}\parallel\mathbf{t}$, black symbols), the magnetic field diminishes the area of the domain with $\boldsymbol{\ell} \parallel \mathbf{t}$ until it completely vanishes. By further increasing the magnetic field magnitude, $\boldsymbol{\ell}$ rotates away from the magnetic field resulting in an increase of $\ell_{j}$ and thus to an increase of $\rho_\mathrm{long}$ according to Eq.~(\ref{eq:averaging_SMR}). 

\section{Results and Discussion}

\begin{figure}[t]
\includegraphics[width=1.0\columnwidth]{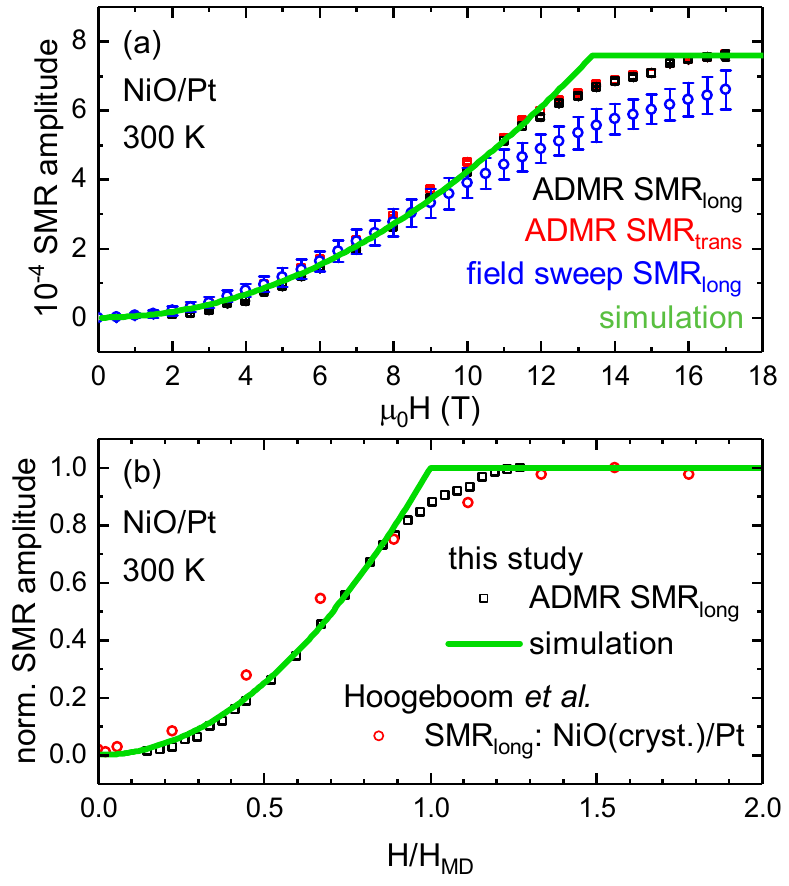}
    \caption{(a) SMR amplitude of the NiO(111)/Pt thin film bilayer obtained from ADMR measurements at 300\,K at different applied magnetic fields (cf.~Fig.~\ref{fig:abb2}) using the longitudinal (black symbols) and transverse (red symbols) resistivities as well as data extracted from field sweep measurements (blue symbols). The data are compared to the analytical model based on a magnetic field induced domain redistribution in NiO (green line). (b) Comparison of the normalized SMR amplitude of our NiO/Pt thin film heterostructure to the data published by Hoogeboom \textit{et al.}\cite{Hoogeboom:2017} measured on a NiO/Pt sample using a NiO single crystal. The magnetic field is normalized to the monodomainization field $H^\mathrm{film}_\mathrm{MD}=13.4$\,T and $H^\mathrm{cryst}_\mathrm{MD}=4.1$\,T, respectively.}
 \label{fig:abb4}
\end{figure}

The SMR amplitudes obtained from the ADMR as well as the field-sweep measurements are depicted in Fig.~\ref{fig:abb4}(a) (black, red, and blue symbols) as a function of the external magnetic field $H$. Almost no difference between $\mathrm{SMR}_\mathrm{long}$ and $\mathrm{SMR}_\mathrm{trans}$ is observable, which is in agreement with the notion $\rho_1 = \rho_3$ in the SMR theory.\cite{Chen:2013,Althammer:2013} Furthermore, the SMR values derived from field sweep measurements at fixed magnetic field orientations $\alpha$ are in good agreement with the SMR amplitudes obtained from the ADMR measurements. The deviation at high magnetic fields is mainly caused by a slight temperature variation during the field-sweep measurement, as well as a small misalignment of the Hall bar with respect to the current direction. As expected from Eq.~(\ref{eq:SMR_amplitudes}), we observe a quadratic dependence of the SMR amplitudes as a function of $H$ for small magnetic fields. At higher fields, the SMR amplitudes start to saturate. 

To determine $H_\mathrm{MD}$ in our sample, we fit the SMR amplitudes according to Eqs.~(\ref{eq:SMR_amplitudes}). The best fit was obtained for $\mu_0 H^\mathrm{film}_\mathrm{MD}=13.4$\,T (cf.~green solid line in Fig.~\ref{fig:abb4}). $H_\mathrm{MD}$ is highly dependent on the specific sample used for the experiment. This is further evidenced by evaluating the data published by Hoogeboom and coworkers using a NiO/Pt sample based on a NiO single crystal.\cite{Hoogeboom:2017} For this sample we derive a significantly lower monodomainization field of $\mu_0 H^\mathrm{cryst}_\mathrm{MD}=4.1$\,T. This demonstrates that our model can generically explain the SMR using antiferromagnets with strong magnetoelastic coupling (see Fig.~\ref{fig:abb4}(b)). 

In addition, the destressing field $H_\mathrm{dest}=H^2_\mathrm{MD}/(4 H_\mathrm{ex})$ can easily be determined via the SMR. By using $\mu_0 H_\mathrm{ex}=968.4$\,T\cite{Machado:2017} and the derived values for $H_\mathrm{MD}$, we obtain $\mu_0 H^\mathrm{film}_\mathrm{dest}=46$\,mT for our NiO/Pt thin film sample and $\mu_0 H^\mathrm{cryst}_\mathrm{dest}=4$\,mT for the bulk NiO/Pt sample of Ref.~\onlinecite{Hoogeboom:2017}. This demonstrates that $H_\mathrm{dest}$, which is a measure of the local internal stress fields created by the antiferromagnetic ordering, is highly sensitive to the details of the sample. In the NiO/Pt thin film heterostructure, $H_\mathrm{dest}$ is one order of magnitude larger than in NiO/Pt hybrids using NiO single crystals. This is mainly caused by the elastic clamping of the NiO thin film on the Al$_2$O$_3$ substrate in case of the NiO/Pt thin film bilayer.

Comparing the simulation to our experimental data, we find a very good agreement for $H < H_\mathrm{MD}$. However, at $H \geq H_\mathrm{MD}$ the experimental data do not show the expected saturation of the SMR amplitude. This is most likely caused by finite pinning effects, which can affect the magnetic structure in two ways. On the one hand, they can pin magnetic domain walls and on the other hand, they can affect the direction $\boldsymbol{\ell}^{(k)}$ within one domain $k$ instead of a combination of the magnetic anisotropy and the Zeeman energy. These pinning effects, which are neglected in the simulation, prevent the formation of a single domain state in the NiO layer within the magnetic field range up to 17\,T.

\section{Conclusion}

In summary, we study the angular dependence of the SMR in thin film NiO/Pt heterostructures, revealing a phase shift of $90^\circ$ when compared to the SMR in YIG/Pt and a pronounced field dependence of the SMR amplitude. We further present a comprehensive model for the SMR effect in multidomain antiferromagnets. Our systematic study of the field dependence of the SMR amplitude and the subsequent comparison to simulations provides conclusive evidence for magnetic field induced domain redistribution due to movable antiferromagnetic domain walls as the dominant effect for the field dependence of the SMR in NiO/Pt heterostructures. We further demonstrate that the SMR is a versatile and simple tool to investigate not only the magnetic spin structure, but also local magnetoelastic effects in antiferromagnetic materials. 

We thank T.~Brenninger, A.~Habel, and K.~Helm-Knapp for technical support and J.~Barker as well as U.~K.~R\"{o}{\ss}ler for fruitful discussions. We gratefully acknowledge financial support by the Deutsche Forschungsgemeinschaft via SPP 1538 (Project No. GO 944/4 and No. GR 1132/18), the German Excellence Initiative via the "Nanosystems Initiative Munich (NIM)", and a Laura-Bassi stipend of the Technical University of Munich. O.G. acknowledges support from the Alexander von Humbolt Foundation, the ERC Synergy Grant SC2 (No. 610115), the Transregional Collaborative Research Center (SFB/TRR) 173 SPIN+X, and EU FET Open RIA Grant No. 766566.

\providecommand{\noopsort}[1]{}\providecommand{\singleletter}[1]{#1}%


\begin{thebibliography}{63}%
\makeatletter
\providecommand \@ifxundefined [1]{%
 \@ifx{#1\undefined}
}%
\providecommand \@ifnum [1]{%
 \ifnum #1\expandafter \@firstoftwo
 \else \expandafter \@secondoftwo
 \fi
}%
\providecommand \@ifx [1]{%
 \ifx #1\expandafter \@firstoftwo
 \else \expandafter \@secondoftwo
 \fi
}%
\providecommand \natexlab [1]{#1}%
\providecommand \enquote  [1]{``#1''}%
\providecommand \bibnamefont  [1]{#1}%
\providecommand \bibfnamefont [1]{#1}%
\providecommand \citenamefont [1]{#1}%
\providecommand \href@noop [0]{\@secondoftwo}%
\providecommand \href [0]{\begingroup \@sanitize@url \@href}%
\providecommand \@href[1]{\@@startlink{#1}\@@href}%
\providecommand \@@href[1]{\endgroup#1\@@endlink}%
\providecommand \@sanitize@url [0]{\catcode `\\12\catcode `\$12\catcode
  `\&12\catcode `\#12\catcode `\^12\catcode `\_12\catcode `\%12\relax}%
\providecommand \@@startlink[1]{}%
\providecommand \@@endlink[0]{}%
\providecommand \url  [0]{\begingroup\@sanitize@url \@url }%
\providecommand \@url [1]{\endgroup\@href {#1}{\urlprefix }}%
\providecommand \urlprefix  [0]{URL }%
\providecommand \Eprint [0]{\href }%
\providecommand \doibase [0]{http://dx.doi.org/}%
\providecommand \selectlanguage [0]{\@gobble}%
\providecommand \bibinfo  [0]{\@secondoftwo}%
\providecommand \bibfield  [0]{\@secondoftwo}%
\providecommand \translation [1]{[#1]}%
\providecommand \BibitemOpen [0]{}%
\providecommand \bibitemStop [0]{}%
\providecommand \bibitemNoStop [0]{.\EOS\space}%
\providecommand \EOS [0]{\spacefactor3000\relax}%
\providecommand \BibitemShut  [1]{\csname bibitem#1\endcsname}%
\let\auto@bib@innerbib\@empty
\bibitem [{\citenamefont {Shick}\ \emph {et~al.}(2010)\citenamefont {Shick},
  \citenamefont {Khmelevskyi}, \citenamefont {Mryasov}, \citenamefont
  {Wunderlich},\ and\ \citenamefont {Jungwirth}}]{Shick:2010}%
  \BibitemOpen
  \bibfield  {author} {\bibinfo {author} {\bibfnamefont {A.~B.}\ \bibnamefont
  {Shick}}, \bibinfo {author} {\bibfnamefont {S.}~\bibnamefont {Khmelevskyi}},
  \bibinfo {author} {\bibfnamefont {O.~N.}\ \bibnamefont {Mryasov}}, \bibinfo
  {author} {\bibfnamefont {J.}~\bibnamefont {Wunderlich}}, \ and\ \bibinfo
  {author} {\bibfnamefont {T.}~\bibnamefont {Jungwirth}},\ }\href {\doibase
  10.1103/PhysRevB.81.212409} {\bibfield  {journal} {\bibinfo  {journal} {Phys.
  Rev. B}\ }\textbf {\bibinfo {volume} {81}},\ \bibinfo {pages} {212409}
  (\bibinfo {year} {2010})}\BibitemShut {NoStop}%
\bibitem [{\citenamefont {MacDonald}\ and\ \citenamefont
  {Tsoi}(2011)}]{MacDonald:2011}%
  \BibitemOpen
  \bibfield  {author} {\bibinfo {author} {\bibfnamefont {A.~H.}\ \bibnamefont
  {MacDonald}}\ and\ \bibinfo {author} {\bibfnamefont {M.}~\bibnamefont
  {Tsoi}},\ }\href {\doibase 10.1098/rsta.2011.0014} {\bibfield  {journal}
  {\bibinfo  {journal} {Philosophical Transactions of the Royal Society of
  London A: Mathematical, Physical and Engineering Sciences}\ }\textbf
  {\bibinfo {volume} {369}},\ \bibinfo {pages} {3098} (\bibinfo {year}
  {2011})}\BibitemShut {NoStop}%
\bibitem [{\citenamefont {Barthem}\ \emph {et~al.}(2013)\citenamefont
  {Barthem}, \citenamefont {Colin}, \citenamefont {Mayaffre}, \citenamefont
  {Julien},\ and\ \citenamefont {Givord}}]{Barthem:2013}%
  \BibitemOpen
  \bibfield  {author} {\bibinfo {author} {\bibfnamefont {V.~M. T.~S.}\
  \bibnamefont {Barthem}}, \bibinfo {author} {\bibfnamefont {C.~V.}\
  \bibnamefont {Colin}}, \bibinfo {author} {\bibfnamefont {H.}~\bibnamefont
  {Mayaffre}}, \bibinfo {author} {\bibfnamefont {M.-H.}\ \bibnamefont
  {Julien}}, \ and\ \bibinfo {author} {\bibfnamefont {D.}~\bibnamefont
  {Givord}},\ }\href {http://dx.doi.org/10.1038/ncomms3892} {\bibfield
  {journal} {\bibinfo  {journal} {Nature Communications}\ }\textbf {\bibinfo
  {volume} {4}},\ \bibinfo {pages} {2892} (\bibinfo {year} {2013})}\BibitemShut
  {NoStop}%
\bibitem [{\citenamefont {\ifmmode~\check{Z}\else \v{Z}\fi{}elezn\'y}\ \emph
  {et~al.}(2014)\citenamefont {\ifmmode~\check{Z}\else \v{Z}\fi{}elezn\'y},
  \citenamefont {Gao}, \citenamefont {V\'yborn\'y}, \citenamefont {Zemen},
  \citenamefont {Ma\ifmmode~\check{s}\else \v{s}\fi{}ek}, \citenamefont
  {Manchon}, \citenamefont {Wunderlich}, \citenamefont {Sinova},\ and\
  \citenamefont {Jungwirth}}]{Zelezny:2014}%
  \BibitemOpen
  \bibfield  {author} {\bibinfo {author} {\bibfnamefont {J.}~\bibnamefont
  {\ifmmode~\check{Z}\else \v{Z}\fi{}elezn\'y}}, \bibinfo {author}
  {\bibfnamefont {H.}~\bibnamefont {Gao}}, \bibinfo {author} {\bibfnamefont
  {K.}~\bibnamefont {V\'yborn\'y}}, \bibinfo {author} {\bibfnamefont
  {J.}~\bibnamefont {Zemen}}, \bibinfo {author} {\bibfnamefont
  {J.}~\bibnamefont {Ma\ifmmode~\check{s}\else \v{s}\fi{}ek}}, \bibinfo
  {author} {\bibfnamefont {A.}~\bibnamefont {Manchon}}, \bibinfo {author}
  {\bibfnamefont {J.}~\bibnamefont {Wunderlich}}, \bibinfo {author}
  {\bibfnamefont {J.}~\bibnamefont {Sinova}}, \ and\ \bibinfo {author}
  {\bibfnamefont {T.}~\bibnamefont {Jungwirth}},\ }\href {\doibase
  10.1103/PhysRevLett.113.157201} {\bibfield  {journal} {\bibinfo  {journal}
  {Phys. Rev. Lett.}\ }\textbf {\bibinfo {volume} {113}},\ \bibinfo {pages}
  {157201} (\bibinfo {year} {2014})}\BibitemShut {NoStop}%
\bibitem [{\citenamefont {Gomonay}\ and\ \citenamefont
  {Loktev}(2014)}]{Gomonay:2014}%
  \BibitemOpen
  \bibfield  {author} {\bibinfo {author} {\bibfnamefont {E.~V.}\ \bibnamefont
  {Gomonay}}\ and\ \bibinfo {author} {\bibfnamefont {V.~M.}\ \bibnamefont
  {Loktev}},\ }\href {\doibase 10.1063/1.4862467} {\bibfield  {journal}
  {\bibinfo  {journal} {Low Temperature Physics}\ }\textbf {\bibinfo {volume}
  {40}},\ \bibinfo {pages} {17} (\bibinfo {year} {2014})},\ \Eprint
  {http://arxiv.org/abs/http://dx.doi.org/10.1063/1.4862467}
  {http://dx.doi.org/10.1063/1.4862467} \BibitemShut {NoStop}%
\bibitem [{\citenamefont {Jungwirth}\ \emph {et~al.}(2016)\citenamefont
  {Jungwirth}, \citenamefont {Marti}, \citenamefont {Wadley},\ and\
  \citenamefont {Wunderlich}}]{Jungwirth:2016}%
  \BibitemOpen
  \bibfield  {author} {\bibinfo {author} {\bibfnamefont {T.}~\bibnamefont
  {Jungwirth}}, \bibinfo {author} {\bibfnamefont {X.}~\bibnamefont {Marti}},
  \bibinfo {author} {\bibfnamefont {P.}~\bibnamefont {Wadley}}, \ and\ \bibinfo
  {author} {\bibfnamefont {J.}~\bibnamefont {Wunderlich}},\ }\href
  {http://dx.doi.org/10.1038/nnano.2016.18} {\bibfield  {journal} {\bibinfo
  {journal} {Nat Nano}\ }\textbf {\bibinfo {volume} {11}},\ \bibinfo {pages}
  {231} (\bibinfo {year} {2016})}\BibitemShut {NoStop}%
\bibitem [{\citenamefont {Baltz}\ \emph {et~al.}()\citenamefont {Baltz},
  \citenamefont {Manchon}, \citenamefont {Tsoi}, \citenamefont {Moriyama},
  \citenamefont {Ono}, ,\ and\ \citenamefont {Tserkovnyak}}]{Baltz:2017}%
  \BibitemOpen
  \bibfield  {author} {\bibinfo {author} {\bibfnamefont {V.}~\bibnamefont
  {Baltz}}, \bibinfo {author} {\bibfnamefont {A.}~\bibnamefont {Manchon}},
  \bibinfo {author} {\bibfnamefont {M.}~\bibnamefont {Tsoi}}, \bibinfo {author}
  {\bibfnamefont {T.}~\bibnamefont {Moriyama}}, \bibinfo {author}
  {\bibfnamefont {T.}~\bibnamefont {Ono}}, , \ and\ \bibinfo {author}
  {\bibfnamefont {Y.}~\bibnamefont {Tserkovnyak}},\ }\href@noop {} {}\bibinfo
  {note} {ArXiv:1606.04284 [Rev. Mod. Phys. (2017) (to be
  published)].}\BibitemShut {Stop}%
\bibitem [{\citenamefont {Marti}\ \emph {et~al.}(2014)\citenamefont {Marti},
  \citenamefont {Fina}, \citenamefont {Frontera}, \citenamefont {Liu},
  \citenamefont {Wadley}, \citenamefont {He}, \citenamefont {Paull},
  \citenamefont {Clarkson}, \citenamefont {Kudrnovsk{\'y}}, \citenamefont
  {Turek}, \citenamefont {Kune{\v s}}, \citenamefont {Yi}, \citenamefont {Chu},
  \citenamefont {Nelson}, \citenamefont {You}, \citenamefont {Arenholz},
  \citenamefont {Salahuddin}, \citenamefont {Fontcuberta}, \citenamefont
  {Jungwirth},\ and\ \citenamefont {Ramesh}}]{Marti:2014}%
  \BibitemOpen
  \bibfield  {author} {\bibinfo {author} {\bibfnamefont {X.}~\bibnamefont
  {Marti}}, \bibinfo {author} {\bibfnamefont {I.}~\bibnamefont {Fina}},
  \bibinfo {author} {\bibfnamefont {C.}~\bibnamefont {Frontera}}, \bibinfo
  {author} {\bibfnamefont {J.}~\bibnamefont {Liu}}, \bibinfo {author}
  {\bibfnamefont {P.}~\bibnamefont {Wadley}}, \bibinfo {author} {\bibfnamefont
  {Q.}~\bibnamefont {He}}, \bibinfo {author} {\bibfnamefont {R.~J.}\
  \bibnamefont {Paull}}, \bibinfo {author} {\bibfnamefont {J.~D.}\ \bibnamefont
  {Clarkson}}, \bibinfo {author} {\bibfnamefont {J.}~\bibnamefont
  {Kudrnovsk{\'y}}}, \bibinfo {author} {\bibfnamefont {I.}~\bibnamefont
  {Turek}}, \bibinfo {author} {\bibfnamefont {J.}~\bibnamefont {Kune{\v s}}},
  \bibinfo {author} {\bibfnamefont {D.}~\bibnamefont {Yi}}, \bibinfo {author}
  {\bibfnamefont {J.-H.}\ \bibnamefont {Chu}}, \bibinfo {author} {\bibfnamefont
  {C.~T.}\ \bibnamefont {Nelson}}, \bibinfo {author} {\bibfnamefont
  {L.}~\bibnamefont {You}}, \bibinfo {author} {\bibfnamefont {E.}~\bibnamefont
  {Arenholz}}, \bibinfo {author} {\bibfnamefont {S.}~\bibnamefont
  {Salahuddin}}, \bibinfo {author} {\bibfnamefont {J.}~\bibnamefont
  {Fontcuberta}}, \bibinfo {author} {\bibfnamefont {T.}~\bibnamefont
  {Jungwirth}}, \ and\ \bibinfo {author} {\bibfnamefont {R.}~\bibnamefont
  {Ramesh}},\ }\href {http://dx.doi.org/10.1038/nmat3861} {\bibfield  {journal}
  {\bibinfo  {journal} {Nat Mater}\ }\textbf {\bibinfo {volume} {13}},\
  \bibinfo {pages} {367} (\bibinfo {year} {2014})}\BibitemShut {NoStop}%
\bibitem [{\citenamefont {Wadley}\ \emph {et~al.}(2016)\citenamefont {Wadley},
  \citenamefont {Howells}, \citenamefont {{\v Z}elezn{\'y}}, \citenamefont
  {Andrews}, \citenamefont {Hills}, \citenamefont {Campion}, \citenamefont
  {Nov{\'a}k}, \citenamefont {Olejn{\'\i}k}, \citenamefont {Maccherozzi},
  \citenamefont {Dhesi}, \citenamefont {Martin}, \citenamefont {Wagner},
  \citenamefont {Wunderlich}, \citenamefont {Freimuth}, \citenamefont
  {Mokrousov}, \citenamefont {Kune{\v s}}, \citenamefont {Chauhan},
  \citenamefont {Grzybowski}, \citenamefont {Rushforth}, \citenamefont
  {Edmonds}, \citenamefont {Gallagher},\ and\ \citenamefont
  {Jungwirth}}]{Wadley:2016}%
  \BibitemOpen
  \bibfield  {author} {\bibinfo {author} {\bibfnamefont {P.}~\bibnamefont
  {Wadley}}, \bibinfo {author} {\bibfnamefont {B.}~\bibnamefont {Howells}},
  \bibinfo {author} {\bibfnamefont {J.}~\bibnamefont {{\v Z}elezn{\'y}}},
  \bibinfo {author} {\bibfnamefont {C.}~\bibnamefont {Andrews}}, \bibinfo
  {author} {\bibfnamefont {V.}~\bibnamefont {Hills}}, \bibinfo {author}
  {\bibfnamefont {R.~P.}\ \bibnamefont {Campion}}, \bibinfo {author}
  {\bibfnamefont {V.}~\bibnamefont {Nov{\'a}k}}, \bibinfo {author}
  {\bibfnamefont {K.}~\bibnamefont {Olejn{\'\i}k}}, \bibinfo {author}
  {\bibfnamefont {F.}~\bibnamefont {Maccherozzi}}, \bibinfo {author}
  {\bibfnamefont {S.~S.}\ \bibnamefont {Dhesi}}, \bibinfo {author}
  {\bibfnamefont {S.~Y.}\ \bibnamefont {Martin}}, \bibinfo {author}
  {\bibfnamefont {T.}~\bibnamefont {Wagner}}, \bibinfo {author} {\bibfnamefont
  {J.}~\bibnamefont {Wunderlich}}, \bibinfo {author} {\bibfnamefont
  {F.}~\bibnamefont {Freimuth}}, \bibinfo {author} {\bibfnamefont
  {Y.}~\bibnamefont {Mokrousov}}, \bibinfo {author} {\bibfnamefont
  {J.}~\bibnamefont {Kune{\v s}}}, \bibinfo {author} {\bibfnamefont {J.~S.}\
  \bibnamefont {Chauhan}}, \bibinfo {author} {\bibfnamefont {M.~J.}\
  \bibnamefont {Grzybowski}}, \bibinfo {author} {\bibfnamefont {A.~W.}\
  \bibnamefont {Rushforth}}, \bibinfo {author} {\bibfnamefont {K.~W.}\
  \bibnamefont {Edmonds}}, \bibinfo {author} {\bibfnamefont {B.~L.}\
  \bibnamefont {Gallagher}}, \ and\ \bibinfo {author} {\bibfnamefont
  {T.}~\bibnamefont {Jungwirth}},\ }\href {\doibase 10.1126/science.aab1031}
  {\bibfield  {journal} {\bibinfo  {journal} {Science}\ }\textbf {\bibinfo
  {volume} {351}},\ \bibinfo {pages} {587} (\bibinfo {year} {2016})},\ \Eprint
  {http://arxiv.org/abs/http://science.sciencemag.org/content/351/6273/587.full.pdf}
  {http://science.sciencemag.org/content/351/6273/587.full.pdf} \BibitemShut
  {NoStop}%
\bibitem [{\citenamefont {Satoh}\ \emph {et~al.}(2014)\citenamefont {Satoh},
  \citenamefont {Iida}, \citenamefont {Higuchi}, \citenamefont {Fiebig},\ and\
  \citenamefont {Shimura}}]{Satoh:2014}%
  \BibitemOpen
  \bibfield  {author} {\bibinfo {author} {\bibfnamefont {T.}~\bibnamefont
  {Satoh}}, \bibinfo {author} {\bibfnamefont {R.}~\bibnamefont {Iida}},
  \bibinfo {author} {\bibfnamefont {T.}~\bibnamefont {Higuchi}}, \bibinfo
  {author} {\bibfnamefont {M.}~\bibnamefont {Fiebig}}, \ and\ \bibinfo {author}
  {\bibfnamefont {T.}~\bibnamefont {Shimura}},\ }\href {\doibase
  10.1038/nphoton.2014.273} {\bibfield  {journal} {\bibinfo  {journal} {Nature
  Photonics}\ }\textbf {\bibinfo {volume} {9}},\ \bibinfo {pages} {2} (\bibinfo
  {year} {2014})}\BibitemShut {NoStop}%
\bibitem [{\citenamefont {Duong}\ \emph {et~al.}(2004)\citenamefont {Duong},
  \citenamefont {Satoh},\ and\ \citenamefont {Fiebig}}]{Duong:2004}%
  \BibitemOpen
  \bibfield  {author} {\bibinfo {author} {\bibfnamefont {N.~P.}\ \bibnamefont
  {Duong}}, \bibinfo {author} {\bibfnamefont {T.}~\bibnamefont {Satoh}}, \ and\
  \bibinfo {author} {\bibfnamefont {M.}~\bibnamefont {Fiebig}},\ }\href
  {\doibase 10.1103/PhysRevLett.93.117402} {\bibfield  {journal} {\bibinfo
  {journal} {Phys. Rev. Lett.}\ }\textbf {\bibinfo {volume} {93}},\ \bibinfo
  {pages} {117402} (\bibinfo {year} {2004})}\BibitemShut {NoStop}%
\bibitem [{\citenamefont {Satoh}\ \emph {et~al.}(2010)\citenamefont {Satoh},
  \citenamefont {Cho}, \citenamefont {Iida}, \citenamefont {Shimura},
  \citenamefont {Kuroda}, \citenamefont {Ueda}, \citenamefont {Ueda},
  \citenamefont {Ivanov}, \citenamefont {Nori},\ and\ \citenamefont
  {Fiebig}}]{Satoh:2010}%
  \BibitemOpen
  \bibfield  {author} {\bibinfo {author} {\bibfnamefont {T.}~\bibnamefont
  {Satoh}}, \bibinfo {author} {\bibfnamefont {S.-J.}\ \bibnamefont {Cho}},
  \bibinfo {author} {\bibfnamefont {R.}~\bibnamefont {Iida}}, \bibinfo {author}
  {\bibfnamefont {T.}~\bibnamefont {Shimura}}, \bibinfo {author} {\bibfnamefont
  {K.}~\bibnamefont {Kuroda}}, \bibinfo {author} {\bibfnamefont
  {H.}~\bibnamefont {Ueda}}, \bibinfo {author} {\bibfnamefont {Y.}~\bibnamefont
  {Ueda}}, \bibinfo {author} {\bibfnamefont {B.~A.}\ \bibnamefont {Ivanov}},
  \bibinfo {author} {\bibfnamefont {F.}~\bibnamefont {Nori}}, \ and\ \bibinfo
  {author} {\bibfnamefont {M.}~\bibnamefont {Fiebig}},\ }\href {\doibase
  10.1103/PhysRevLett.105.077402} {\bibfield  {journal} {\bibinfo  {journal}
  {Phys. Rev. Lett.}\ }\textbf {\bibinfo {volume} {105}},\ \bibinfo {pages}
  {077402} (\bibinfo {year} {2010})}\BibitemShut {NoStop}%
\bibitem [{\citenamefont {Kampfrath}\ \emph {et~al.}(2010)\citenamefont
  {Kampfrath}, \citenamefont {Sell}, \citenamefont {Klatt}, \citenamefont
  {Pashkin}, \citenamefont {M\"{a}hrlein}, \citenamefont {Dekorsy},
  \citenamefont {Wolf}, \citenamefont {Fiebig}, \citenamefont {Leitenstorfer},\
  and\ \citenamefont {Huber}}]{Kampfrath:2010}%
  \BibitemOpen
  \bibfield  {author} {\bibinfo {author} {\bibfnamefont {T.}~\bibnamefont
  {Kampfrath}}, \bibinfo {author} {\bibfnamefont {A.}~\bibnamefont {Sell}},
  \bibinfo {author} {\bibfnamefont {G.}~\bibnamefont {Klatt}}, \bibinfo
  {author} {\bibfnamefont {A.}~\bibnamefont {Pashkin}}, \bibinfo {author}
  {\bibfnamefont {S.}~\bibnamefont {M\"{a}hrlein}}, \bibinfo {author}
  {\bibfnamefont {T.}~\bibnamefont {Dekorsy}}, \bibinfo {author} {\bibfnamefont
  {M.}~\bibnamefont {Wolf}}, \bibinfo {author} {\bibfnamefont {M.}~\bibnamefont
  {Fiebig}}, \bibinfo {author} {\bibfnamefont {A.}~\bibnamefont
  {Leitenstorfer}}, \ and\ \bibinfo {author} {\bibfnamefont {R.}~\bibnamefont
  {Huber}},\ }\href@noop {} {\bibfield  {journal} {\bibinfo  {journal} {Nature
  Photonics}\ }\textbf {\bibinfo {volume} {5}},\ \bibinfo {pages} {31}
  (\bibinfo {year} {2010})}\BibitemShut {NoStop}%
\bibitem [{\citenamefont {Chen}\ \emph {et~al.}(2014)\citenamefont {Chen},
  \citenamefont {Niu},\ and\ \citenamefont {MacDonald}}]{Chen:2014}%
  \BibitemOpen
  \bibfield  {author} {\bibinfo {author} {\bibfnamefont {H.}~\bibnamefont
  {Chen}}, \bibinfo {author} {\bibfnamefont {Q.}~\bibnamefont {Niu}}, \ and\
  \bibinfo {author} {\bibfnamefont {A.~H.}\ \bibnamefont {MacDonald}},\ }\href
  {\doibase 10.1103/PhysRevLett.112.017205} {\bibfield  {journal} {\bibinfo
  {journal} {Phys. Rev. Lett.}\ }\textbf {\bibinfo {volume} {112}},\ \bibinfo
  {pages} {017205} (\bibinfo {year} {2014})}\BibitemShut {NoStop}%
\bibitem [{\citenamefont {Mendes}\ \emph {et~al.}(2014)\citenamefont {Mendes},
  \citenamefont {Cunha}, \citenamefont {Alves~Santos}, \citenamefont {Ribeiro},
  \citenamefont {Machado}, \citenamefont {Rodr\'{\i}guez-Su\'arez},
  \citenamefont {Azevedo},\ and\ \citenamefont {Rezende}}]{Mendes:2014}%
  \BibitemOpen
  \bibfield  {author} {\bibinfo {author} {\bibfnamefont {J.~B.~S.}\
  \bibnamefont {Mendes}}, \bibinfo {author} {\bibfnamefont {R.~O.}\
  \bibnamefont {Cunha}}, \bibinfo {author} {\bibfnamefont {O.}~\bibnamefont
  {Alves~Santos}}, \bibinfo {author} {\bibfnamefont {P.~R.~T.}\ \bibnamefont
  {Ribeiro}}, \bibinfo {author} {\bibfnamefont {F.~L.~A.}\ \bibnamefont
  {Machado}}, \bibinfo {author} {\bibfnamefont {R.~L.}\ \bibnamefont
  {Rodr\'{\i}guez-Su\'arez}}, \bibinfo {author} {\bibfnamefont
  {A.}~\bibnamefont {Azevedo}}, \ and\ \bibinfo {author} {\bibfnamefont
  {S.~M.}\ \bibnamefont {Rezende}},\ }\href {\doibase
  10.1103/PhysRevB.89.140406} {\bibfield  {journal} {\bibinfo  {journal} {Phys.
  Rev. B}\ }\textbf {\bibinfo {volume} {89}},\ \bibinfo {pages} {140406}
  (\bibinfo {year} {2014})}\BibitemShut {NoStop}%
\bibitem [{\citenamefont {Zhang}\ \emph {et~al.}(2014)\citenamefont {Zhang},
  \citenamefont {Jungfleisch}, \citenamefont {Jiang}, \citenamefont {Pearson},
  \citenamefont {Hoffmann}, \citenamefont {Freimuth},\ and\ \citenamefont
  {Mokrousov}}]{Zhang:2014}%
  \BibitemOpen
  \bibfield  {author} {\bibinfo {author} {\bibfnamefont {W.}~\bibnamefont
  {Zhang}}, \bibinfo {author} {\bibfnamefont {M.~B.}\ \bibnamefont
  {Jungfleisch}}, \bibinfo {author} {\bibfnamefont {W.}~\bibnamefont {Jiang}},
  \bibinfo {author} {\bibfnamefont {J.~E.}\ \bibnamefont {Pearson}}, \bibinfo
  {author} {\bibfnamefont {A.}~\bibnamefont {Hoffmann}}, \bibinfo {author}
  {\bibfnamefont {F.}~\bibnamefont {Freimuth}}, \ and\ \bibinfo {author}
  {\bibfnamefont {Y.}~\bibnamefont {Mokrousov}},\ }\href {\doibase
  10.1103/PhysRevLett.113.196602} {\bibfield  {journal} {\bibinfo  {journal}
  {Phys. Rev. Lett.}\ }\textbf {\bibinfo {volume} {113}},\ \bibinfo {pages}
  {196602} (\bibinfo {year} {2014})}\BibitemShut {NoStop}%
\bibitem [{\citenamefont {Ou}\ \emph {et~al.}(2016)\citenamefont {Ou},
  \citenamefont {Shi}, \citenamefont {Ralph},\ and\ \citenamefont
  {Buhrman}}]{Ou:2016}%
  \BibitemOpen
  \bibfield  {author} {\bibinfo {author} {\bibfnamefont {Y.}~\bibnamefont
  {Ou}}, \bibinfo {author} {\bibfnamefont {S.}~\bibnamefont {Shi}}, \bibinfo
  {author} {\bibfnamefont {D.~C.}\ \bibnamefont {Ralph}}, \ and\ \bibinfo
  {author} {\bibfnamefont {R.~A.}\ \bibnamefont {Buhrman}},\ }\href {\doibase
  10.1103/PhysRevB.93.220405} {\bibfield  {journal} {\bibinfo  {journal} {Phys.
  Rev. B}\ }\textbf {\bibinfo {volume} {93}},\ \bibinfo {pages} {220405}
  (\bibinfo {year} {2016})}\BibitemShut {NoStop}%
\bibitem [{\citenamefont {Ohnuma}\ \emph {et~al.}(2013)\citenamefont {Ohnuma},
  \citenamefont {Adachi}, \citenamefont {Saitoh},\ and\ \citenamefont
  {Maekawa}}]{Ohnuma:2013}%
  \BibitemOpen
  \bibfield  {author} {\bibinfo {author} {\bibfnamefont {Y.}~\bibnamefont
  {Ohnuma}}, \bibinfo {author} {\bibfnamefont {H.}~\bibnamefont {Adachi}},
  \bibinfo {author} {\bibfnamefont {E.}~\bibnamefont {Saitoh}}, \ and\ \bibinfo
  {author} {\bibfnamefont {S.}~\bibnamefont {Maekawa}},\ }\href {\doibase
  10.1103/PhysRevB.87.014423} {\bibfield  {journal} {\bibinfo  {journal} {Phys.
  Rev. B}\ }\textbf {\bibinfo {volume} {87}},\ \bibinfo {pages} {014423}
  (\bibinfo {year} {2013})}\BibitemShut {NoStop}%
\bibitem [{\citenamefont {Seki}\ \emph {et~al.}(2015)\citenamefont {Seki},
  \citenamefont {Ideue}, \citenamefont {Kubota}, \citenamefont {Kozuka},
  \citenamefont {Takagi}, \citenamefont {Nakamura}, \citenamefont {Kaneko},
  \citenamefont {Kawasaki},\ and\ \citenamefont {Tokura}}]{Seki:2015}%
  \BibitemOpen
  \bibfield  {author} {\bibinfo {author} {\bibfnamefont {S.}~\bibnamefont
  {Seki}}, \bibinfo {author} {\bibfnamefont {T.}~\bibnamefont {Ideue}},
  \bibinfo {author} {\bibfnamefont {M.}~\bibnamefont {Kubota}}, \bibinfo
  {author} {\bibfnamefont {Y.}~\bibnamefont {Kozuka}}, \bibinfo {author}
  {\bibfnamefont {R.}~\bibnamefont {Takagi}}, \bibinfo {author} {\bibfnamefont
  {M.}~\bibnamefont {Nakamura}}, \bibinfo {author} {\bibfnamefont
  {Y.}~\bibnamefont {Kaneko}}, \bibinfo {author} {\bibfnamefont
  {M.}~\bibnamefont {Kawasaki}}, \ and\ \bibinfo {author} {\bibfnamefont
  {Y.}~\bibnamefont {Tokura}},\ }\href {\doibase
  10.1103/PhysRevLett.115.266601} {\bibfield  {journal} {\bibinfo  {journal}
  {Phys. Rev. Lett.}\ }\textbf {\bibinfo {volume} {115}},\ \bibinfo {pages}
  {266601} (\bibinfo {year} {2015})}\BibitemShut {NoStop}%
\bibitem [{\citenamefont {Rezende}\ \emph {et~al.}(2016)\citenamefont
  {Rezende}, \citenamefont {Rodr\'{\i}guez-Su\'arez},\ and\ \citenamefont
  {Azevedo}}]{Rezende:2016}%
  \BibitemOpen
  \bibfield  {author} {\bibinfo {author} {\bibfnamefont {S.~M.}\ \bibnamefont
  {Rezende}}, \bibinfo {author} {\bibfnamefont {R.~L.}\ \bibnamefont
  {Rodr\'{\i}guez-Su\'arez}}, \ and\ \bibinfo {author} {\bibfnamefont
  {A.}~\bibnamefont {Azevedo}},\ }\href {\doibase 10.1103/PhysRevB.93.014425}
  {\bibfield  {journal} {\bibinfo  {journal} {Phys. Rev. B}\ }\textbf {\bibinfo
  {volume} {93}},\ \bibinfo {pages} {014425} (\bibinfo {year}
  {2016})}\BibitemShut {NoStop}%
\bibitem [{\citenamefont {Wu}\ \emph {et~al.}(2016)\citenamefont {Wu},
  \citenamefont {Zhang}, \citenamefont {KC}, \citenamefont {Borisov},
  \citenamefont {Pearson}, \citenamefont {Jiang}, \citenamefont {Lederman},
  \citenamefont {Hoffmann},\ and\ \citenamefont {Bhattacharya}}]{Wu:2016a}%
  \BibitemOpen
  \bibfield  {author} {\bibinfo {author} {\bibfnamefont {S.~M.}\ \bibnamefont
  {Wu}}, \bibinfo {author} {\bibfnamefont {W.}~\bibnamefont {Zhang}}, \bibinfo
  {author} {\bibfnamefont {A.}~\bibnamefont {KC}}, \bibinfo {author}
  {\bibfnamefont {P.}~\bibnamefont {Borisov}}, \bibinfo {author} {\bibfnamefont
  {J.~E.}\ \bibnamefont {Pearson}}, \bibinfo {author} {\bibfnamefont {J.~S.}\
  \bibnamefont {Jiang}}, \bibinfo {author} {\bibfnamefont {D.}~\bibnamefont
  {Lederman}}, \bibinfo {author} {\bibfnamefont {A.}~\bibnamefont {Hoffmann}},
  \ and\ \bibinfo {author} {\bibfnamefont {A.}~\bibnamefont {Bhattacharya}},\
  }\href {\doibase 10.1103/PhysRevLett.116.097204} {\bibfield  {journal}
  {\bibinfo  {journal} {Phys. Rev. Lett.}\ }\textbf {\bibinfo {volume} {116}},\
  \bibinfo {pages} {097204} (\bibinfo {year} {2016})}\BibitemShut {NoStop}%
\bibitem [{\citenamefont {Cheng}\ \emph {et~al.}(2016)\citenamefont {Cheng},
  \citenamefont {Okamoto},\ and\ \citenamefont {Xiao}}]{Cheng:2016}%
  \BibitemOpen
  \bibfield  {author} {\bibinfo {author} {\bibfnamefont {R.}~\bibnamefont
  {Cheng}}, \bibinfo {author} {\bibfnamefont {S.}~\bibnamefont {Okamoto}}, \
  and\ \bibinfo {author} {\bibfnamefont {D.}~\bibnamefont {Xiao}},\ }\href
  {\doibase 10.1103/PhysRevLett.117.217202} {\bibfield  {journal} {\bibinfo
  {journal} {Phys. Rev. Lett.}\ }\textbf {\bibinfo {volume} {117}},\ \bibinfo
  {pages} {217202} (\bibinfo {year} {2016})}\BibitemShut {NoStop}%
\bibitem [{\citenamefont {Wang}\ \emph {et~al.}(2014)\citenamefont {Wang},
  \citenamefont {Du}, \citenamefont {Hammel},\ and\ \citenamefont
  {Yang}}]{Wang:2014}%
  \BibitemOpen
  \bibfield  {author} {\bibinfo {author} {\bibfnamefont {H.}~\bibnamefont
  {Wang}}, \bibinfo {author} {\bibfnamefont {C.}~\bibnamefont {Du}}, \bibinfo
  {author} {\bibfnamefont {P.~C.}\ \bibnamefont {Hammel}}, \ and\ \bibinfo
  {author} {\bibfnamefont {F.}~\bibnamefont {Yang}},\ }\href {\doibase
  10.1103/PhysRevLett.113.097202} {\bibfield  {journal} {\bibinfo  {journal}
  {Phys. Rev. Lett.}\ }\textbf {\bibinfo {volume} {113}},\ \bibinfo {pages}
  {097202} (\bibinfo {year} {2014})}\BibitemShut {NoStop}%
\bibitem [{\citenamefont {Hahn}\ \emph {et~al.}(2014)\citenamefont {Hahn},
  \citenamefont {de~Loubens}, \citenamefont {Naletov}, \citenamefont {Youssef},
  \citenamefont {Klein},\ and\ \citenamefont {Viret}}]{Hahn:2014}%
  \BibitemOpen
  \bibfield  {author} {\bibinfo {author} {\bibfnamefont {C.}~\bibnamefont
  {Hahn}}, \bibinfo {author} {\bibfnamefont {G.}~\bibnamefont {de~Loubens}},
  \bibinfo {author} {\bibfnamefont {V.~V.}\ \bibnamefont {Naletov}}, \bibinfo
  {author} {\bibfnamefont {J.}~\bibnamefont {Ben Youssef}}, \bibinfo {author}
  {\bibfnamefont {O.}~\bibnamefont {Klein}}, \ and\ \bibinfo {author}
  {\bibfnamefont {M.}~\bibnamefont {Viret}},\ }\href
  {http://stacks.iop.org/0295-5075/108/i=5/a=57005} {\bibfield  {journal}
  {\bibinfo  {journal} {EPL (Europhysics Letters)}\ }\textbf {\bibinfo {volume}
  {108}},\ \bibinfo {pages} {57005} (\bibinfo {year} {2014})}\BibitemShut
  {NoStop}%
\bibitem [{\citenamefont {Moriyama}\ \emph {et~al.}(2015)\citenamefont
  {Moriyama}, \citenamefont {Takei}, \citenamefont {Nagata}, \citenamefont
  {Yoshimura}, \citenamefont {Matsuzaki}, \citenamefont {Terashima},
  \citenamefont {Tserkovnyak},\ and\ \citenamefont {Ono}}]{Moriyama:2015}%
  \BibitemOpen
  \bibfield  {author} {\bibinfo {author} {\bibfnamefont {T.}~\bibnamefont
  {Moriyama}}, \bibinfo {author} {\bibfnamefont {S.}~\bibnamefont {Takei}},
  \bibinfo {author} {\bibfnamefont {M.}~\bibnamefont {Nagata}}, \bibinfo
  {author} {\bibfnamefont {Y.}~\bibnamefont {Yoshimura}}, \bibinfo {author}
  {\bibfnamefont {N.}~\bibnamefont {Matsuzaki}}, \bibinfo {author}
  {\bibfnamefont {T.}~\bibnamefont {Terashima}}, \bibinfo {author}
  {\bibfnamefont {Y.}~\bibnamefont {Tserkovnyak}}, \ and\ \bibinfo {author}
  {\bibfnamefont {T.}~\bibnamefont {Ono}},\ }\href {\doibase 10.1063/1.4918990}
  {\bibfield  {journal} {\bibinfo  {journal} {Applied Physics Letters}\
  }\textbf {\bibinfo {volume} {106}},\ \bibinfo {pages} {162406} (\bibinfo
  {year} {2015})},\ \Eprint
  {http://arxiv.org/abs/http://dx.doi.org/10.1063/1.4918990}
  {http://dx.doi.org/10.1063/1.4918990} \BibitemShut {NoStop}%
\bibitem [{\citenamefont {Lin}\ \emph {et~al.}(2016)\citenamefont {Lin},
  \citenamefont {Chen}, \citenamefont {Zhang},\ and\ \citenamefont
  {Chien}}]{Lin:2016}%
  \BibitemOpen
  \bibfield  {author} {\bibinfo {author} {\bibfnamefont {W.}~\bibnamefont
  {Lin}}, \bibinfo {author} {\bibfnamefont {K.}~\bibnamefont {Chen}}, \bibinfo
  {author} {\bibfnamefont {S.}~\bibnamefont {Zhang}}, \ and\ \bibinfo {author}
  {\bibfnamefont {C.~L.}\ \bibnamefont {Chien}},\ }\href {\doibase
  10.1103/PhysRevLett.116.186601} {\bibfield  {journal} {\bibinfo  {journal}
  {Phys. Rev. Lett.}\ }\textbf {\bibinfo {volume} {116}},\ \bibinfo {pages}
  {186601} (\bibinfo {year} {2016})}\BibitemShut {NoStop}%
\bibitem [{\citenamefont {Shang}\ \emph {et~al.}(2016)\citenamefont {Shang},
  \citenamefont {Zhan}, \citenamefont {Yang}, \citenamefont {Zuo},
  \citenamefont {Xie}, \citenamefont {Liu}, \citenamefont {Zhang},
  \citenamefont {Zhang}, \citenamefont {Li}, \citenamefont {Wang},
  \citenamefont {Wu}, \citenamefont {Zhang},\ and\ \citenamefont
  {Li}}]{Shang:2016}%
  \BibitemOpen
  \bibfield  {author} {\bibinfo {author} {\bibfnamefont {T.}~\bibnamefont
  {Shang}}, \bibinfo {author} {\bibfnamefont {Q.~F.}\ \bibnamefont {Zhan}},
  \bibinfo {author} {\bibfnamefont {H.~L.}\ \bibnamefont {Yang}}, \bibinfo
  {author} {\bibfnamefont {Z.~H.}\ \bibnamefont {Zuo}}, \bibinfo {author}
  {\bibfnamefont {Y.~L.}\ \bibnamefont {Xie}}, \bibinfo {author} {\bibfnamefont
  {L.~P.}\ \bibnamefont {Liu}}, \bibinfo {author} {\bibfnamefont {S.~L.}\
  \bibnamefont {Zhang}}, \bibinfo {author} {\bibfnamefont {Y.}~\bibnamefont
  {Zhang}}, \bibinfo {author} {\bibfnamefont {H.~H.}\ \bibnamefont {Li}},
  \bibinfo {author} {\bibfnamefont {B.~M.}\ \bibnamefont {Wang}}, \bibinfo
  {author} {\bibfnamefont {Y.~H.}\ \bibnamefont {Wu}}, \bibinfo {author}
  {\bibfnamefont {S.}~\bibnamefont {Zhang}}, \ and\ \bibinfo {author}
  {\bibfnamefont {R.-W.}\ \bibnamefont {Li}},\ }\href {\doibase
  10.1063/1.4959573} {\bibfield  {journal} {\bibinfo  {journal} {Applied
  Physics Letters}\ }\textbf {\bibinfo {volume} {109}},\ \bibinfo {pages}
  {032410} (\bibinfo {year} {2016})},\ \Eprint
  {http://arxiv.org/abs/http://dx.doi.org/10.1063/1.4959573}
  {http://dx.doi.org/10.1063/1.4959573} \BibitemShut {NoStop}%
\bibitem [{\citenamefont {Qaiumzadeh}\ \emph {et~al.}(2017)\citenamefont
  {Qaiumzadeh}, \citenamefont {Skarsv\aa{}g}, \citenamefont {Holmqvist},\ and\
  \citenamefont {Brataas}}]{Qaiumzadeh:2017}%
  \BibitemOpen
  \bibfield  {author} {\bibinfo {author} {\bibfnamefont {A.}~\bibnamefont
  {Qaiumzadeh}}, \bibinfo {author} {\bibfnamefont {H.}~\bibnamefont
  {Skarsv\aa{}g}}, \bibinfo {author} {\bibfnamefont {C.}~\bibnamefont
  {Holmqvist}}, \ and\ \bibinfo {author} {\bibfnamefont {A.}~\bibnamefont
  {Brataas}},\ }\href {\doibase 10.1103/PhysRevLett.118.137201} {\bibfield
  {journal} {\bibinfo  {journal} {Phys. Rev. Lett.}\ }\textbf {\bibinfo
  {volume} {118}},\ \bibinfo {pages} {137201} (\bibinfo {year}
  {2017})}\BibitemShut {NoStop}%
\bibitem [{\citenamefont {Wang}\ \emph {et~al.}(2015)\citenamefont {Wang},
  \citenamefont {Du}, \citenamefont {Hammel},\ and\ \citenamefont
  {Yang}}]{Wang:2015}%
  \BibitemOpen
  \bibfield  {author} {\bibinfo {author} {\bibfnamefont {H.}~\bibnamefont
  {Wang}}, \bibinfo {author} {\bibfnamefont {C.}~\bibnamefont {Du}}, \bibinfo
  {author} {\bibfnamefont {P.~C.}\ \bibnamefont {Hammel}}, \ and\ \bibinfo
  {author} {\bibfnamefont {F.}~\bibnamefont {Yang}},\ }\href {\doibase
  10.1103/PhysRevB.91.220410} {\bibfield  {journal} {\bibinfo  {journal} {Phys.
  Rev. B}\ }\textbf {\bibinfo {volume} {91}},\ \bibinfo {pages} {220410}
  (\bibinfo {year} {2015})}\BibitemShut {NoStop}%
\bibitem [{\citenamefont {Hou}\ \emph {et~al.}(2017)\citenamefont {Hou},
  \citenamefont {Qiu}, \citenamefont {Barker}, \citenamefont {Sato},
  \citenamefont {Yamamoto}, \citenamefont {V\'elez}, \citenamefont
  {Gomez-Perez}, \citenamefont {Hueso}, \citenamefont {Casanova},\ and\
  \citenamefont {Saitoh}}]{Hou:2017}%
  \BibitemOpen
  \bibfield  {author} {\bibinfo {author} {\bibfnamefont {D.}~\bibnamefont
  {Hou}}, \bibinfo {author} {\bibfnamefont {Z.}~\bibnamefont {Qiu}}, \bibinfo
  {author} {\bibfnamefont {J.}~\bibnamefont {Barker}}, \bibinfo {author}
  {\bibfnamefont {K.}~\bibnamefont {Sato}}, \bibinfo {author} {\bibfnamefont
  {K.}~\bibnamefont {Yamamoto}}, \bibinfo {author} {\bibfnamefont
  {S.}~\bibnamefont {V\'elez}}, \bibinfo {author} {\bibfnamefont {J.~M.}\
  \bibnamefont {Gomez-Perez}}, \bibinfo {author} {\bibfnamefont {L.~E.}\
  \bibnamefont {Hueso}}, \bibinfo {author} {\bibfnamefont {F.}~\bibnamefont
  {Casanova}}, \ and\ \bibinfo {author} {\bibfnamefont {E.}~\bibnamefont
  {Saitoh}},\ }\href {\doibase 10.1103/PhysRevLett.118.147202} {\bibfield
  {journal} {\bibinfo  {journal} {Phys. Rev. Lett.}\ }\textbf {\bibinfo
  {volume} {118}},\ \bibinfo {pages} {147202} (\bibinfo {year}
  {2017})}\BibitemShut {NoStop}%
\bibitem [{\citenamefont {Lin}\ and\ \citenamefont {Chien}(2017)}]{Lin:2017}%
  \BibitemOpen
  \bibfield  {author} {\bibinfo {author} {\bibfnamefont {W.}~\bibnamefont
  {Lin}}\ and\ \bibinfo {author} {\bibfnamefont {C.~L.}\ \bibnamefont
  {Chien}},\ }\href {\doibase 10.1103/PhysRevLett.118.067202} {\bibfield
  {journal} {\bibinfo  {journal} {Phys. Rev. Lett.}\ }\textbf {\bibinfo
  {volume} {118}},\ \bibinfo {pages} {067202} (\bibinfo {year}
  {2017})}\BibitemShut {NoStop}%
\bibitem [{\citenamefont {Hung}\ \emph {et~al.}(2017)\citenamefont {Hung},
  \citenamefont {Hahn}, \citenamefont {Chang}, \citenamefont {Wu},
  \citenamefont {Ohldag},\ and\ \citenamefont {Kent}}]{Hung:2017}%
  \BibitemOpen
  \bibfield  {author} {\bibinfo {author} {\bibfnamefont {Y.-M.}\ \bibnamefont
  {Hung}}, \bibinfo {author} {\bibfnamefont {C.}~\bibnamefont {Hahn}}, \bibinfo
  {author} {\bibfnamefont {H.}~\bibnamefont {Chang}}, \bibinfo {author}
  {\bibfnamefont {M.}~\bibnamefont {Wu}}, \bibinfo {author} {\bibfnamefont
  {H.}~\bibnamefont {Ohldag}}, \ and\ \bibinfo {author} {\bibfnamefont {A.~D.}\
  \bibnamefont {Kent}},\ }\href {\doibase 10.1063/1.4972998} {\bibfield
  {journal} {\bibinfo  {journal} {AIP Advances}\ }\textbf {\bibinfo {volume}
  {7}},\ \bibinfo {pages} {055903} (\bibinfo {year} {2017})},\ \Eprint
  {http://arxiv.org/abs/http://dx.doi.org/10.1063/1.4972998}
  {http://dx.doi.org/10.1063/1.4972998} \BibitemShut {NoStop}%
\bibitem [{\citenamefont {Nakayama}\ \emph {et~al.}(2013)\citenamefont
  {Nakayama}, \citenamefont {Althammer}, \citenamefont {Chen}, \citenamefont
  {Uchida}, \citenamefont {Kajiwara}, \citenamefont {Kikuchi}, \citenamefont
  {Ohtani}, \citenamefont {Gepr\"ags}, \citenamefont {Opel}, \citenamefont
  {Takahashi}, \citenamefont {Gross}, \citenamefont {Bauer}, \citenamefont
  {Goennenwein},\ and\ \citenamefont {Saitoh}}]{Nakayama:2013}%
  \BibitemOpen
  \bibfield  {author} {\bibinfo {author} {\bibfnamefont {H.}~\bibnamefont
  {Nakayama}}, \bibinfo {author} {\bibfnamefont {M.}~\bibnamefont {Althammer}},
  \bibinfo {author} {\bibfnamefont {Y.-T.}\ \bibnamefont {Chen}}, \bibinfo
  {author} {\bibfnamefont {K.}~\bibnamefont {Uchida}}, \bibinfo {author}
  {\bibfnamefont {Y.}~\bibnamefont {Kajiwara}}, \bibinfo {author}
  {\bibfnamefont {D.}~\bibnamefont {Kikuchi}}, \bibinfo {author} {\bibfnamefont
  {T.}~\bibnamefont {Ohtani}}, \bibinfo {author} {\bibfnamefont
  {S.}~\bibnamefont {Gepr\"ags}}, \bibinfo {author} {\bibfnamefont
  {M.}~\bibnamefont {Opel}}, \bibinfo {author} {\bibfnamefont {S.}~\bibnamefont
  {Takahashi}}, \bibinfo {author} {\bibfnamefont {R.}~\bibnamefont {Gross}},
  \bibinfo {author} {\bibfnamefont {G.~E.~W.}\ \bibnamefont {Bauer}}, \bibinfo
  {author} {\bibfnamefont {S.~T.~B.}\ \bibnamefont {Goennenwein}}, \ and\
  \bibinfo {author} {\bibfnamefont {E.}~\bibnamefont {Saitoh}},\ }\href
  {\doibase 10.1103/PhysRevLett.110.206601} {\bibfield  {journal} {\bibinfo
  {journal} {Phys. Rev. Lett.}\ }\textbf {\bibinfo {volume} {110}},\ \bibinfo
  {pages} {206601} (\bibinfo {year} {2013})}\BibitemShut {NoStop}%
\bibitem [{\citenamefont {Althammer}\ \emph {et~al.}(2013)\citenamefont
  {Althammer}, \citenamefont {Meyer}, \citenamefont {Nakayama}, \citenamefont
  {Schreier}, \citenamefont {Altmannshofer}, \citenamefont {Weiler},
  \citenamefont {Huebl}, \citenamefont {Gepr\"ags}, \citenamefont {Opel},
  \citenamefont {Gross}, \citenamefont {Meier}, \citenamefont {Klewe},
  \citenamefont {Kuschel}, \citenamefont {Schmalhorst}, \citenamefont {Reiss},
  \citenamefont {Shen}, \citenamefont {Gupta}, \citenamefont {Chen},
  \citenamefont {Bauer}, \citenamefont {Saitoh},\ and\ \citenamefont
  {Goennenwein}}]{Althammer:2013}%
  \BibitemOpen
  \bibfield  {author} {\bibinfo {author} {\bibfnamefont {M.}~\bibnamefont
  {Althammer}}, \bibinfo {author} {\bibfnamefont {S.}~\bibnamefont {Meyer}},
  \bibinfo {author} {\bibfnamefont {H.}~\bibnamefont {Nakayama}}, \bibinfo
  {author} {\bibfnamefont {M.}~\bibnamefont {Schreier}}, \bibinfo {author}
  {\bibfnamefont {S.}~\bibnamefont {Altmannshofer}}, \bibinfo {author}
  {\bibfnamefont {M.}~\bibnamefont {Weiler}}, \bibinfo {author} {\bibfnamefont
  {H.}~\bibnamefont {Huebl}}, \bibinfo {author} {\bibfnamefont
  {S.}~\bibnamefont {Gepr\"ags}}, \bibinfo {author} {\bibfnamefont
  {M.}~\bibnamefont {Opel}}, \bibinfo {author} {\bibfnamefont {R.}~\bibnamefont
  {Gross}}, \bibinfo {author} {\bibfnamefont {D.}~\bibnamefont {Meier}},
  \bibinfo {author} {\bibfnamefont {C.}~\bibnamefont {Klewe}}, \bibinfo
  {author} {\bibfnamefont {T.}~\bibnamefont {Kuschel}}, \bibinfo {author}
  {\bibfnamefont {J.-M.}\ \bibnamefont {Schmalhorst}}, \bibinfo {author}
  {\bibfnamefont {G.}~\bibnamefont {Reiss}}, \bibinfo {author} {\bibfnamefont
  {L.}~\bibnamefont {Shen}}, \bibinfo {author} {\bibfnamefont {A.}~\bibnamefont
  {Gupta}}, \bibinfo {author} {\bibfnamefont {Y.-T.}\ \bibnamefont {Chen}},
  \bibinfo {author} {\bibfnamefont {G.~E.~W.}\ \bibnamefont {Bauer}}, \bibinfo
  {author} {\bibfnamefont {E.}~\bibnamefont {Saitoh}}, \ and\ \bibinfo {author}
  {\bibfnamefont {S.~T.~B.}\ \bibnamefont {Goennenwein}},\ }\href {\doibase
  10.1103/PhysRevB.87.224401} {\bibfield  {journal} {\bibinfo  {journal} {Phys.
  Rev. B}\ }\textbf {\bibinfo {volume} {87}},\ \bibinfo {pages} {224401}
  (\bibinfo {year} {2013})}\BibitemShut {NoStop}%
\bibitem [{\citenamefont {Vlietstra}\ \emph {et~al.}(2013)\citenamefont
  {Vlietstra}, \citenamefont {Shan}, \citenamefont {Castel}, \citenamefont {van
  Wees},\ and\ \citenamefont {Ben~Youssef}}]{Vlietstra:2013}%
  \BibitemOpen
  \bibfield  {author} {\bibinfo {author} {\bibfnamefont {N.}~\bibnamefont
  {Vlietstra}}, \bibinfo {author} {\bibfnamefont {J.}~\bibnamefont {Shan}},
  \bibinfo {author} {\bibfnamefont {V.}~\bibnamefont {Castel}}, \bibinfo
  {author} {\bibfnamefont {B.~J.}\ \bibnamefont {van Wees}}, \ and\ \bibinfo
  {author} {\bibfnamefont {J.}~\bibnamefont {Ben~Youssef}},\ }\href {\doibase
  10.1103/PhysRevB.87.184421} {\bibfield  {journal} {\bibinfo  {journal} {Phys.
  Rev. B}\ }\textbf {\bibinfo {volume} {87}},\ \bibinfo {pages} {184421}
  (\bibinfo {year} {2013})}\BibitemShut {NoStop}%
\bibitem [{\citenamefont {Hirsch}(1999)}]{Hirsch:1999}%
  \BibitemOpen
  \bibfield  {author} {\bibinfo {author} {\bibfnamefont {J.~E.}\ \bibnamefont
  {Hirsch}},\ }\href {\doibase 10.1103/PhysRevLett.83.1834} {\bibfield
  {journal} {\bibinfo  {journal} {Phys. Rev. Lett.}\ }\textbf {\bibinfo
  {volume} {83}},\ \bibinfo {pages} {1834} (\bibinfo {year}
  {1999})}\BibitemShut {NoStop}%
\bibitem [{\citenamefont {Chen}\ \emph {et~al.}(2013)\citenamefont {Chen},
  \citenamefont {Takahashi}, \citenamefont {Nakayama}, \citenamefont
  {Althammer}, \citenamefont {Goennenwein}, \citenamefont {Saitoh},\ and\
  \citenamefont {Bauer}}]{Chen:2013}%
  \BibitemOpen
  \bibfield  {author} {\bibinfo {author} {\bibfnamefont {Y.-T.}\ \bibnamefont
  {Chen}}, \bibinfo {author} {\bibfnamefont {S.}~\bibnamefont {Takahashi}},
  \bibinfo {author} {\bibfnamefont {H.}~\bibnamefont {Nakayama}}, \bibinfo
  {author} {\bibfnamefont {M.}~\bibnamefont {Althammer}}, \bibinfo {author}
  {\bibfnamefont {S.~T.~B.}\ \bibnamefont {Goennenwein}}, \bibinfo {author}
  {\bibfnamefont {E.}~\bibnamefont {Saitoh}}, \ and\ \bibinfo {author}
  {\bibfnamefont {G.~E.~W.}\ \bibnamefont {Bauer}},\ }\href {\doibase
  10.1103/PhysRevB.87.144411} {\bibfield  {journal} {\bibinfo  {journal} {Phys.
  Rev. B}\ }\textbf {\bibinfo {volume} {87}},\ \bibinfo {pages} {144411}
  (\bibinfo {year} {2013})}\BibitemShut {NoStop}%
\bibitem [{\citenamefont {Hahn}\ \emph {et~al.}(2013)\citenamefont {Hahn},
  \citenamefont {de~Loubens}, \citenamefont {Klein}, \citenamefont {Viret},
  \citenamefont {Naletov},\ and\ \citenamefont {Ben~Youssef}}]{Hahn:2013}%
  \BibitemOpen
  \bibfield  {author} {\bibinfo {author} {\bibfnamefont {C.}~\bibnamefont
  {Hahn}}, \bibinfo {author} {\bibfnamefont {G.}~\bibnamefont {de~Loubens}},
  \bibinfo {author} {\bibfnamefont {O.}~\bibnamefont {Klein}}, \bibinfo
  {author} {\bibfnamefont {M.}~\bibnamefont {Viret}}, \bibinfo {author}
  {\bibfnamefont {V.~V.}\ \bibnamefont {Naletov}}, \ and\ \bibinfo {author}
  {\bibfnamefont {J.}~\bibnamefont {Ben~Youssef}},\ }\href {\doibase
  10.1103/PhysRevB.87.174417} {\bibfield  {journal} {\bibinfo  {journal} {Phys.
  Rev. B}\ }\textbf {\bibinfo {volume} {87}},\ \bibinfo {pages} {174417}
  (\bibinfo {year} {2013})}\BibitemShut {NoStop}%
\bibitem [{\citenamefont {Marmion}\ \emph {et~al.}(2014)\citenamefont
  {Marmion}, \citenamefont {Ali}, \citenamefont {McLaren}, \citenamefont
  {Williams},\ and\ \citenamefont {Hickey}}]{Marmion:2014}%
  \BibitemOpen
  \bibfield  {author} {\bibinfo {author} {\bibfnamefont {S.~R.}\ \bibnamefont
  {Marmion}}, \bibinfo {author} {\bibfnamefont {M.}~\bibnamefont {Ali}},
  \bibinfo {author} {\bibfnamefont {M.}~\bibnamefont {McLaren}}, \bibinfo
  {author} {\bibfnamefont {D.~A.}\ \bibnamefont {Williams}}, \ and\ \bibinfo
  {author} {\bibfnamefont {B.~J.}\ \bibnamefont {Hickey}},\ }\href {\doibase
  10.1103/PhysRevB.89.220404} {\bibfield  {journal} {\bibinfo  {journal} {Phys.
  Rev. B}\ }\textbf {\bibinfo {volume} {89}},\ \bibinfo {pages} {220404}
  (\bibinfo {year} {2014})}\BibitemShut {NoStop}%
\bibitem [{\citenamefont {Meyer}\ \emph {et~al.}(2014)\citenamefont {Meyer},
  \citenamefont {Althammer}, \citenamefont {Gepr\"{a}gs}, \citenamefont {Opel},
  \citenamefont {Gross},\ and\ \citenamefont {Goennenwein}}]{Meyer:2014}%
  \BibitemOpen
  \bibfield  {author} {\bibinfo {author} {\bibfnamefont {S.}~\bibnamefont
  {Meyer}}, \bibinfo {author} {\bibfnamefont {M.}~\bibnamefont {Althammer}},
  \bibinfo {author} {\bibfnamefont {S.}~\bibnamefont {Gepr\"{a}gs}}, \bibinfo
  {author} {\bibfnamefont {M.}~\bibnamefont {Opel}}, \bibinfo {author}
  {\bibfnamefont {R.}~\bibnamefont {Gross}}, \ and\ \bibinfo {author}
  {\bibfnamefont {S.~T.~B.}\ \bibnamefont {Goennenwein}},\ }\href {\doibase
  10.1063/1.4885086} {\bibfield  {journal} {\bibinfo  {journal} {Applied
  Physics Letters}\ }\textbf {\bibinfo {volume} {104}},\ \bibinfo {eid}
  {242411} (\bibinfo {year} {2014}),\ 10.1063/1.4885086}\BibitemShut {NoStop}%
\bibitem [{\citenamefont {Aldosary}\ \emph {et~al.}(2016)\citenamefont
  {Aldosary}, \citenamefont {Li}, \citenamefont {Tang}, \citenamefont {Xu},
  \citenamefont {Zheng}, \citenamefont {Bozhilov},\ and\ \citenamefont
  {Shi}}]{Aldosary:2016}%
  \BibitemOpen
  \bibfield  {author} {\bibinfo {author} {\bibfnamefont {M.}~\bibnamefont
  {Aldosary}}, \bibinfo {author} {\bibfnamefont {J.}~\bibnamefont {Li}},
  \bibinfo {author} {\bibfnamefont {C.}~\bibnamefont {Tang}}, \bibinfo {author}
  {\bibfnamefont {Y.}~\bibnamefont {Xu}}, \bibinfo {author} {\bibfnamefont
  {J.-G.}\ \bibnamefont {Zheng}}, \bibinfo {author} {\bibfnamefont {K.~N.}\
  \bibnamefont {Bozhilov}}, \ and\ \bibinfo {author} {\bibfnamefont
  {J.}~\bibnamefont {Shi}},\ }\href {\doibase 10.1063/1.4953454} {\bibfield
  {journal} {\bibinfo  {journal} {Applied Physics Letters}\ }\textbf {\bibinfo
  {volume} {108}},\ \bibinfo {eid} {242401} (\bibinfo {year} {2016}),\
  10.1063/1.4953454}\BibitemShut {NoStop}%
\bibitem [{\citenamefont {Isasa}\ \emph {et~al.}(2014)\citenamefont {Isasa},
  \citenamefont {Bedoya-Pinto}, \citenamefont {Vélez}, \citenamefont {Golmar},
  \citenamefont {Sánchez}, \citenamefont {Hueso}, \citenamefont
  {Fontcuberta},\ and\ \citenamefont {Casanova}}]{Isasa:2014}%
  \BibitemOpen
  \bibfield  {author} {\bibinfo {author} {\bibfnamefont {M.}~\bibnamefont
  {Isasa}}, \bibinfo {author} {\bibfnamefont {A.}~\bibnamefont {Bedoya-Pinto}},
  \bibinfo {author} {\bibfnamefont {S.}~\bibnamefont {Vélez}}, \bibinfo
  {author} {\bibfnamefont {F.}~\bibnamefont {Golmar}}, \bibinfo {author}
  {\bibfnamefont {F.}~\bibnamefont {Sánchez}}, \bibinfo {author}
  {\bibfnamefont {L.~E.}\ \bibnamefont {Hueso}}, \bibinfo {author}
  {\bibfnamefont {J.}~\bibnamefont {Fontcuberta}}, \ and\ \bibinfo {author}
  {\bibfnamefont {F.}~\bibnamefont {Casanova}},\ }\href {\doibase
  10.1063/1.4897544} {\bibfield  {journal} {\bibinfo  {journal} {Applied
  Physics Letters}\ }\textbf {\bibinfo {volume} {105}},\ \bibinfo {eid}
  {142402} (\bibinfo {year} {2014}),\ 10.1063/1.4897544}\BibitemShut {NoStop}%
\bibitem [{\citenamefont {Ganzhorn}\ \emph {et~al.}(2016)\citenamefont
  {Ganzhorn}, \citenamefont {Barker}, \citenamefont {Schlitz}, \citenamefont
  {Piot}, \citenamefont {Ollefs}, \citenamefont {Guillou}, \citenamefont
  {Wilhelm}, \citenamefont {Rogalev}, \citenamefont {Opel}, \citenamefont
  {Althammer}, \citenamefont {Gepr\"ags}, \citenamefont {Huebl}, \citenamefont
  {Gross}, \citenamefont {Bauer},\ and\ \citenamefont
  {Goennenwein}}]{Ganzhorn:2016}%
  \BibitemOpen
  \bibfield  {author} {\bibinfo {author} {\bibfnamefont {K.}~\bibnamefont
  {Ganzhorn}}, \bibinfo {author} {\bibfnamefont {J.}~\bibnamefont {Barker}},
  \bibinfo {author} {\bibfnamefont {R.}~\bibnamefont {Schlitz}}, \bibinfo
  {author} {\bibfnamefont {B.~A.}\ \bibnamefont {Piot}}, \bibinfo {author}
  {\bibfnamefont {K.}~\bibnamefont {Ollefs}}, \bibinfo {author} {\bibfnamefont
  {F.}~\bibnamefont {Guillou}}, \bibinfo {author} {\bibfnamefont
  {F.}~\bibnamefont {Wilhelm}}, \bibinfo {author} {\bibfnamefont
  {A.}~\bibnamefont {Rogalev}}, \bibinfo {author} {\bibfnamefont
  {M.}~\bibnamefont {Opel}}, \bibinfo {author} {\bibfnamefont {M.}~\bibnamefont
  {Althammer}}, \bibinfo {author} {\bibfnamefont {S.}~\bibnamefont
  {Gepr\"ags}}, \bibinfo {author} {\bibfnamefont {H.}~\bibnamefont {Huebl}},
  \bibinfo {author} {\bibfnamefont {R.}~\bibnamefont {Gross}}, \bibinfo
  {author} {\bibfnamefont {G.~E.~W.}\ \bibnamefont {Bauer}}, \ and\ \bibinfo
  {author} {\bibfnamefont {S.~T.~B.}\ \bibnamefont {Goennenwein}},\ }\href
  {\doibase 10.1103/PhysRevB.94.094401} {\bibfield  {journal} {\bibinfo
  {journal} {Phys. Rev. B}\ }\textbf {\bibinfo {volume} {94}},\ \bibinfo
  {pages} {094401} (\bibinfo {year} {2016})}\BibitemShut {NoStop}%
\bibitem [{\citenamefont {Aqeel}\ \emph {et~al.}(2016)\citenamefont {Aqeel},
  \citenamefont {Vlietstra}, \citenamefont {Roy}, \citenamefont {Mostovoy},
  \citenamefont {van Wees},\ and\ \citenamefont {Palstra}}]{Aqeel:2016}%
  \BibitemOpen
  \bibfield  {author} {\bibinfo {author} {\bibfnamefont {A.}~\bibnamefont
  {Aqeel}}, \bibinfo {author} {\bibfnamefont {N.}~\bibnamefont {Vlietstra}},
  \bibinfo {author} {\bibfnamefont {A.}~\bibnamefont {Roy}}, \bibinfo {author}
  {\bibfnamefont {M.}~\bibnamefont {Mostovoy}}, \bibinfo {author}
  {\bibfnamefont {B.~J.}\ \bibnamefont {van Wees}}, \ and\ \bibinfo {author}
  {\bibfnamefont {T.~T.~M.}\ \bibnamefont {Palstra}},\ }\href {\doibase
  10.1103/PhysRevB.94.134418} {\bibfield  {journal} {\bibinfo  {journal} {Phys.
  Rev. B}\ }\textbf {\bibinfo {volume} {94}},\ \bibinfo {pages} {134418}
  (\bibinfo {year} {2016})}\BibitemShut {NoStop}%
\bibitem [{\citenamefont {Ji}\ \emph {et~al.}(2017)\citenamefont {Ji},
  \citenamefont {Miao}, \citenamefont {Meng}, \citenamefont {Ren},
  \citenamefont {Dong}, \citenamefont {Xu}, \citenamefont {Wu},\ and\
  \citenamefont {Jiang}}]{Ji:2017}%
  \BibitemOpen
  \bibfield  {author} {\bibinfo {author} {\bibfnamefont {Y.}~\bibnamefont
  {Ji}}, \bibinfo {author} {\bibfnamefont {J.}~\bibnamefont {Miao}}, \bibinfo
  {author} {\bibfnamefont {K.~K.}\ \bibnamefont {Meng}}, \bibinfo {author}
  {\bibfnamefont {Z.~Y.}\ \bibnamefont {Ren}}, \bibinfo {author} {\bibfnamefont
  {B.~W.}\ \bibnamefont {Dong}}, \bibinfo {author} {\bibfnamefont {X.~G.}\
  \bibnamefont {Xu}}, \bibinfo {author} {\bibfnamefont {Y.}~\bibnamefont {Wu}},
  \ and\ \bibinfo {author} {\bibfnamefont {Y.}~\bibnamefont {Jiang}},\ }\href
  {\doibase 10.1063/1.4989680} {\bibfield  {journal} {\bibinfo  {journal}
  {Appl. Phys. Lett.}\ }\textbf {\bibinfo {volume} {110}},\ \bibinfo {pages}
  {262401} (\bibinfo {year} {2017})}\BibitemShut {NoStop}%
\bibitem [{\citenamefont {Hoogeboom}\ \emph {et~al.}(2017)\citenamefont
  {Hoogeboom}, \citenamefont {Aqeel}, \citenamefont {Kuschel}, \citenamefont
  {Palstra},\ and\ \citenamefont {van Wees}}]{Hoogeboom:2017}%
  \BibitemOpen
  \bibfield  {author} {\bibinfo {author} {\bibfnamefont {G.~R.}\ \bibnamefont
  {Hoogeboom}}, \bibinfo {author} {\bibfnamefont {A.}~\bibnamefont {Aqeel}},
  \bibinfo {author} {\bibfnamefont {T.}~\bibnamefont {Kuschel}}, \bibinfo
  {author} {\bibfnamefont {T.~T.~M.}\ \bibnamefont {Palstra}}, \ and\ \bibinfo
  {author} {\bibfnamefont {B.~J.}\ \bibnamefont {van Wees}},\ }\href {\doibase
  10.1063/1.4997588} {\bibfield  {journal} {\bibinfo  {journal} {Appl. Phys.
  Lett.}\ }\textbf {\bibinfo {volume} {111}},\ \bibinfo {pages} {052409}
  (\bibinfo {year} {2017})}\BibitemShut {NoStop}%

\bibitem [{\citenamefont {{Baldrati}}\ \emph {et~al.}(2017)\citenamefont
  {{Baldrati}}, \citenamefont {{Ross}}, \citenamefont {{Niizeki}},
  \citenamefont {{Schneider}}, \citenamefont {{Ramos}}, \citenamefont {{Cramer}}, \citenamefont {{Gomonay}},
  \citenamefont {{Filianina}}, \citenamefont {{Savchenko}}, \citenamefont {{Heinze}}, \citenamefont {{Kleibert}}, 
	\citenamefont {{Saitoh}}, \citenamefont {{Sinova}},\ and\ \citenamefont
  {{Kl{\"a}ui}}}]{Baldrati:2017}%
  \BibitemOpen
  \bibfield  {author} {
	\bibinfo {author} {\bibfnamefont {L.}~\bibnamefont {{Baldrati}}}, 
	\bibinfo {author} {\bibfnamefont {A.}~\bibnamefont {{Ross}}},
  \bibinfo {author} {\bibfnamefont {T.}~\bibnamefont {{Niizeki}}}, 
	\bibinfo {author} {\bibfnamefont {C.}~\bibnamefont {{Schneider}}}, 
	\bibinfo {author} {\bibfnamefont {R.}~\bibnamefont {{Ramos}}}, 
	\bibinfo {author} {\bibfnamefont {J.}~\bibnamefont {{Cramer}}}, 
	\bibinfo {author} {\bibfnamefont {O.}~\bibnamefont {{Gomonay}}}, 
	\bibinfo {author} {\bibfnamefont {M.}~\bibnamefont {{Filianina}}}, 
	\bibinfo {author} {\bibfnamefont {T.}~\bibnamefont {{Savchenko}}}, 
	\bibinfo {author} {\bibfnamefont {D.}~\bibnamefont {{Heinze}}}, 
	\bibinfo {author} {\bibfnamefont {A.}~\bibnamefont {{Kleibert}}}, 
	\bibinfo {author} {\bibfnamefont {E.}~\bibnamefont {{Saitoh}}}, 
	\bibinfo {author} {\bibfnamefont {J.}~\bibnamefont {{Sinova}}}, \ and\ 
	\bibinfo {author} {\bibfnamefont {M.}~\bibnamefont {{Kl{\"a}ui}}},\ }\href@noop {} {\bibfield
  {journal} {\bibinfo  {journal} {ArXiv e-prints}\ } (\bibinfo {year}
  {2017})},\ \Eprint {http://arxiv.org/abs/1709.00910v3} {arXiv:1709.00910v3
  [cond-mat.mtrl-sci]} \BibitemShut {NoStop}%
\bibitem [{\citenamefont {Wang}\ \emph {et~al.}(2017)\citenamefont {Wang},
  \citenamefont {Hou}, \citenamefont {Qiu}, \citenamefont {Kikkawa},
  \citenamefont {Saitoh},\ and\ \citenamefont {Jin}}]{Wang:2017}%
  \BibitemOpen
  \bibfield  {author} {\bibinfo {author} {\bibfnamefont {H.}~\bibnamefont
  {Wang}}, \bibinfo {author} {\bibfnamefont {D.}~\bibnamefont {Hou}}, \bibinfo
  {author} {\bibfnamefont {Z.}~\bibnamefont {Qiu}}, \bibinfo {author}
  {\bibfnamefont {T.}~\bibnamefont {Kikkawa}}, \bibinfo {author} {\bibfnamefont
  {E.}~\bibnamefont {Saitoh}}, \ and\ \bibinfo {author} {\bibfnamefont
  {X.}~\bibnamefont {Jin}},\ }\href {\doibase 10.1063/1.4986372} {\bibfield
  {journal} {\bibinfo  {journal} {J. Appl. Phys.}\ }\textbf {\bibinfo {volume}
  {122}},\ \bibinfo {pages} {083907} (\bibinfo {year} {2017})}\BibitemShut
  {NoStop}%
\bibitem [{Note1()}]{Note1}%
  \BibitemOpen
  \bibinfo {note} {The
  magnon accumulation at the MI/HM interface created by the spin accumulation
  is usually neglected in the description of the SMR.\cite
  {Cornelissen:2015,Goennenwein:2015}}\BibitemShut {NoStop}%
\bibitem [{Note2()}]{Note2}%
  \BibitemOpen
  \bibinfo {note} {Since
  the expected maximum change of the resistivity of the HM layer via the SMR is
  of the order $10^{-3}$, the difference of the resistivity between a series
  connection and a more complicated resistance network is expected to be
  small.}\BibitemShut {NoStop}%
\bibitem [{Note3()}]{Note3}%
  \BibitemOpen
  \bibinfo {note} {The external field $\protect \mathbf {H}$ also induces
  canted magnetization states leading to a finite net magnetization. Due to the usually 
	large exchange field in antiferromagnets, this canting effect can be
  neglected (exchange approximation), assuming an angle of $180^\circ $ between
  the sublattice magnetizations, i.e. $\protect \mathbf {m}_1 = -\protect
  \mathbf {m}_2$, in the considered magnetic field range $H$.}\BibitemShut
  {NoStop}%
\bibitem [{\citenamefont {Gomonay}\ and\ \citenamefont
  {Loktev}(2002)}]{Gomonay:2002}%
  \BibitemOpen
  \bibfield  {author} {\bibinfo {author} {\bibfnamefont {H.}~\bibnamefont
  {Gomonay}}\ and\ \bibinfo {author} {\bibfnamefont {V.~M.}\ \bibnamefont
  {Loktev}},\ }\href {http://stacks.iop.org/0953-8984/14/i=15/a=310} {\bibfield
   {journal} {\bibinfo  {journal} {J. Phys.: Condens. Matter}\ }\textbf
  {\bibinfo {volume} {14}},\ \bibinfo {pages} {3959} (\bibinfo {year}
  {2002})}\BibitemShut {NoStop}%
\bibitem [{\citenamefont {Gomonay}\ and\ \citenamefont
  {Loktev}(2004)}]{Gomonay:2004}%
  \BibitemOpen
  \bibfield  {author} {\bibinfo {author} {\bibfnamefont {E.~V.}\ \bibnamefont
  {Gomonay}}\ and\ \bibinfo {author} {\bibfnamefont {V.~M.}\ \bibnamefont
  {Loktev}},\ }\href {\doibase 10.1063/1.1808199} {\bibfield  {journal}
  {\bibinfo  {journal} {Low Temperature Physics}\ }\textbf {\bibinfo {volume}
  {30}},\ \bibinfo {pages} {804} (\bibinfo {year} {2004})}\BibitemShut
  {NoStop}%
\bibitem [{\citenamefont {Uchida}\ \emph {et~al.}(1967)\citenamefont {Uchida},
  \citenamefont {Fukuoka}, \citenamefont {Kondoh}, \citenamefont {Takeda},
  \citenamefont {Nakazumi},\ and\ \citenamefont {Nagamiya}}]{Uchida:1967}%
  \BibitemOpen
  \bibfield  {author} {\bibinfo {author} {\bibfnamefont {E.}~\bibnamefont
  {Uchida}}, \bibinfo {author} {\bibfnamefont {N.}~\bibnamefont {Fukuoka}},
  \bibinfo {author} {\bibfnamefont {H.}~\bibnamefont {Kondoh}}, \bibinfo
  {author} {\bibfnamefont {T.}~\bibnamefont {Takeda}}, \bibinfo {author}
  {\bibfnamefont {Y.}~\bibnamefont {Nakazumi}}, \ and\ \bibinfo {author}
  {\bibfnamefont {T.}~\bibnamefont {Nagamiya}},\ }\href {\doibase
  10.1143/JPSJ.23.1197} {\bibfield  {journal} {\bibinfo  {journal} {Journal of
  the Physical Society of Japan}\ }\textbf {\bibinfo {volume} {23}},\ \bibinfo
  {pages} {1197} (\bibinfo {year} {1967})}\BibitemShut {NoStop}%
\bibitem [{\citenamefont {Srinivasan}\ and\ \citenamefont
  {Seehra}(1984)}]{Srinivasan:1984}%
  \BibitemOpen
  \bibfield  {author} {\bibinfo {author} {\bibfnamefont {G.}~\bibnamefont
  {Srinivasan}}\ and\ \bibinfo {author} {\bibfnamefont {M.~S.}\ \bibnamefont
  {Seehra}},\ }\href {\doibase 10.1103/PhysRevB.29.6295} {\bibfield  {journal}
  {\bibinfo  {journal} {Phys. Rev. B}\ }\textbf {\bibinfo {volume} {29}},\
  \bibinfo {pages} {6295} (\bibinfo {year} {1984})}\BibitemShut {NoStop}%
\bibitem [{\citenamefont {Roth}(1958)}]{Roth:1958}%
  \BibitemOpen
  \bibfield  {author} {\bibinfo {author} {\bibfnamefont {W.~L.}\ \bibnamefont
  {Roth}},\ }\href {\doibase 10.1103/PhysRev.110.1333} {\bibfield  {journal}
  {\bibinfo  {journal} {Phys. Rev.}\ }\textbf {\bibinfo {volume} {110}},\
  \bibinfo {pages} {1333} (\bibinfo {year} {1958})}\BibitemShut {NoStop}%
\bibitem [{\citenamefont {Hutchings}\ and\ \citenamefont
  {Samuelsen}(1972)}]{Hutchings:1972}%
  \BibitemOpen
  \bibfield  {author} {\bibinfo {author} {\bibfnamefont {M.~T.}\ \bibnamefont
  {Hutchings}}\ and\ \bibinfo {author} {\bibfnamefont {E.~J.}\ \bibnamefont
  {Samuelsen}},\ }\href {\doibase 10.1103/PhysRevB.6.3447} {\bibfield
  {journal} {\bibinfo  {journal} {Phys. Rev. B}\ }\textbf {\bibinfo {volume}
  {6}},\ \bibinfo {pages} {3447} (\bibinfo {year} {1972})}\BibitemShut
  {NoStop}%
\bibitem [{Note5()}]{Note5}%
  \BibitemOpen
  \bibinfo {note} {The three physically distinguishable antiferromagnetic
  domains result from the threefold symmetry of the NiO(111) plane, i.e. $\protect
  \mathbf {m}_1$ and $\protect \mathbf {m}_2$ are aligned either parallel or
  antiparallel to one of the three $[11\protect \overline {2}]$ easy
  axes.}\BibitemShut {NoStop}%
\bibitem [{\citenamefont {Bartel}\ and\ \citenamefont
  {Morosin}(1971)}]{Bartel:1971}%
  \BibitemOpen
  \bibfield  {author} {\bibinfo {author} {\bibfnamefont {L.~C.}\ \bibnamefont
  {Bartel}}\ and\ \bibinfo {author} {\bibfnamefont {B.}~\bibnamefont
  {Morosin}},\ }\href {\doibase 10.1103/PhysRevB.3.1039} {\bibfield  {journal}
  {\bibinfo  {journal} {Phys. Rev. B}\ }\textbf {\bibinfo {volume} {3}},\
  \bibinfo {pages} {1039} (\bibinfo {year} {1971})}\BibitemShut {NoStop}%
\bibitem [{\citenamefont {Machado}\ \emph {et~al.}(2017)\citenamefont
  {Machado}, \citenamefont {Ribeiro}, \citenamefont {Holanda}, \citenamefont
  {Rodr\'{\i}guez-Su\'arez}, \citenamefont {Azevedo},\ and\ \citenamefont
  {Rezende}}]{Machado:2017}%
  \BibitemOpen
  \bibfield  {author} {\bibinfo {author} {\bibfnamefont {F.~L.~A.}\
  \bibnamefont {Machado}}, \bibinfo {author} {\bibfnamefont {P.~R.~T.}\
  \bibnamefont {Ribeiro}}, \bibinfo {author} {\bibfnamefont {J.}~\bibnamefont
  {Holanda}}, \bibinfo {author} {\bibfnamefont {R.~L.}\ \bibnamefont
  {Rodr\'{\i}guez-Su\'arez}}, \bibinfo {author} {\bibfnamefont
  {A.}~\bibnamefont {Azevedo}}, \ and\ \bibinfo {author} {\bibfnamefont
  {S.~M.}\ \bibnamefont {Rezende}},\ }\href {\doibase
  10.1103/PhysRevB.95.104418} {\bibfield  {journal} {\bibinfo  {journal} {Phys.
  Rev. B}\ }\textbf {\bibinfo {volume} {95}},\ \bibinfo {pages} {104418}
  (\bibinfo {year} {2017})}\BibitemShut {NoStop}%
\bibitem [{\citenamefont {Cornelissen}\ \emph {et~al.}(2015)\citenamefont
  {Cornelissen}, \citenamefont {Liu}, \citenamefont {Duine}, \citenamefont
  {Youssef},\ and\ \citenamefont {van Wees}}]{Cornelissen:2015}%
  \BibitemOpen
  \bibfield  {author} {\bibinfo {author} {\bibfnamefont {L.~J.}\ \bibnamefont
  {Cornelissen}}, \bibinfo {author} {\bibfnamefont {J.}~\bibnamefont {Liu}},
  \bibinfo {author} {\bibfnamefont {R.~A.}\ \bibnamefont {Duine}}, \bibinfo
  {author} {\bibfnamefont {J.}\ \bibnamefont {Ben~Youssef}}, \ and\ \bibinfo
  {author} {\bibfnamefont {B.~J.}\ \bibnamefont {van Wees}},\ }\href
  {http://dx.doi.org/10.1038/nphys3465} {\bibfield  {journal} {\bibinfo
  {journal} {Nat. Phys.}\ }\textbf {\bibinfo {volume} {11}},\ \bibinfo {pages}
  {1022} (\bibinfo {year} {2015})}\BibitemShut {NoStop}%
\bibitem [{\citenamefont {Goennenwein}\ \emph {et~al.}(2015)\citenamefont
  {Goennenwein}, \citenamefont {Schlitz}, \citenamefont {Pernpeintner},
  \citenamefont {Ganzhorn}, \citenamefont {Althammer}, \citenamefont {Gross},\
  and\ \citenamefont {Huebl}}]{Goennenwein:2015}%
  \BibitemOpen
  \bibfield  {author} {\bibinfo {author} {\bibfnamefont {S.~T.~B.}\
  \bibnamefont {Goennenwein}}, \bibinfo {author} {\bibfnamefont
  {R.}~\bibnamefont {Schlitz}}, \bibinfo {author} {\bibfnamefont
  {M.}~\bibnamefont {Pernpeintner}}, \bibinfo {author} {\bibfnamefont
  {K.}~\bibnamefont {Ganzhorn}}, \bibinfo {author} {\bibfnamefont
  {M.}~\bibnamefont {Althammer}}, \bibinfo {author} {\bibfnamefont
  {R.}~\bibnamefont {Gross}}, \ and\ \bibinfo {author} {\bibfnamefont
  {H.}~\bibnamefont {Huebl}},\ }\href {\doibase 10.1063/1.4935074} {\bibfield
  {journal} {\bibinfo  {journal} {Applied Physics Letters}\ }\textbf {\bibinfo
  {volume} {107}},\ \bibinfo {eid} {172405} (\bibinfo {year} {2015}),\
  10.1063/1.4935074}\BibitemShut {NoStop}%
\end{thebibliography}
\end{document}